# Metaheuristics optimized feedforward neural networks for efficient stock price prediction

Bradley J. Pillay[1] and Absalom E. Ezugwu[2]

[1]School of Mathematics, Statistics and Computer Science, University of KwaZulu-Natal, Westville Campus, Private Bag X54001, Durban 4000, South Africa.
[2]School of Computer Science, University of KwaZulu-Natal, King Edward Road, Pietermaritzburg Campus, Pietermaritzburg, KwaZulu-Natal 3201, South Africa.

Corresponding author email: ezugwua@ukzn.ac.za

**Abstract.** The prediction of stock prices is an important task in economics, investment and making financial decisions. This has, for decades, spurred the interest of many researchers to make focused contributions to the design of accurate stock price predictive models; of which some have been utilized to predict the next day opening and closing prices of the stock indices. This paper proposes the design and implementation of a hybrid symbiotic organisms search trained feedforward neural network model for effective and accurate stock price prediction. The symbiotic organisms search algorithm is used as an efficient optimization technique to train the feedforward neural networks, while the resulting training process is used to build a better stock price prediction model. Furthermore, the study also presents a comparative performance evaluation of three different stock price forecasting models; namely, the particle swarm optimization trained feedforward neural network model, the genetic algorithm trained feedforward neural network model and the well-known ARIMA model. The system developed in support of this study utilizes sixteen stock indices as time series datasets for training and testing purpose. Three statistical evaluation measures are used to compare the results of the implemented models, namely the root mean squared error, the mean absolute percentage error and the mean absolution deviation. The computational results obtained reveal that the symbiotic organisms search trained feedforward neural network model exhibits outstanding predictive performance compared to the other models. However, the performance study shows that the three metaheuristics trained feedforward neural network models have promising predictive competence for solving problems of high dimensional nonlinear time series data, which are difficult to capture by traditional models.

## 1. Introduction

The stock market is generally considered very significant in the development of the economy of a country. It provides a form of time series data that is, on the one hand, highly non-linear and non-stationary. Stock market prediction, on the other hand, is the act of trying to determine the future value of a company stock or other financial instruments traded on an exchange. The successful prediction of a stock's future price serves as a guide for investors in relation to their investments. The efficient-market hypothesis suggests that stock prices reflect all currently available information and any price changes that are not based on newly revealed information are inherently unpredictable [1]. Other schools of thought disagree and these individuals rely on various methods and technologies to gain insight on future price information [2]. It is important to note that the perceived value of a stock is often influenced by the earnings per share, the company's book value, the price earnings ratio and dividends per share. Although these factors are the fundamental units influencing the base stock price, the market also exerts some power over a specific stock price at any point in time. This is due to the constant pull and push of demand and supply within the market; this fluctuation may be due to a trader's personal preference, events portrayed by the media, strategic approaches to the stock exchange or perceptions based on other traders' behaviour. Although such fluctuations may be estimated to some extent, based on past behavioural patterns of a particular stock, random events that result in the stock behaving out of its norm are very difficult to predict. Nevertheless, these occurrences are what experienced traders look for in maximising their profits. As such, any insight into these anomalies would be highly valuable to any trader within the market [3].

Traditionally, stock price forecasting has been carried out using time series analysis methods [4]. Linear models, such as the auto regressive integrated moving average (ARIMA) [8, 27] have been used for decades in stock market forecasting. One important challenge with using such linear models is that, they work only for a particular time series data, which means that the model is data-oriented and problem specific. Therefore, the model cannot be used to handle the complex dynamism associated with stock market predictions. It has been noticed that with the emergence of artificial neural networks (ANNs), stock market analysis can now be effectively performed with higher levels of accuracy and accountability for unconceived variables [5, 6, 7]. Some of the advantages of forecasting with ANNs is their ability to easily learn non-linear complex models and handle data with high volatility [24]. However, because ANNs use gradient-based learning techniques, which are influenced by the dynamics of stopping criteria, longer training periods, varying weights due to complex architecture and easy entrapment into local minima, the performance of network models is therefore somewhat limited [25, 26]. The literature on implementation results shows that non-hybridized time series models are outperformed by either improved or hybridized variants of the ANN and ARIMA models [8]. Therefore, the results of these enhancement techniques clearly show that the aforementioned limitation of ANN model can be overcome through hybridization techniques.

The task of training an ANN is considered one of the major challenges in developing a prediction model that is based on ANN. Therefore, the literature abounds with instances of the efficiency and accuracy of ANN-based models having been improved by optimizing their training and learning rate using different global optimization metaheuristic algorithms. For examples, Panapakidis et al. [30] presented a hybrid prediction model for the day-ahead natural gas demand forecasting, based on a combination of multiple techniques, in which a genetic algorithm (GA) was used to optimize the weight vectors of the neural network model, while Fattah et al. [31] proposed a hybrid GA with models based on feedforward neural networks (FFNNs) for automatic text summarization. Many other applications of particle swarm optimization (PSO) for FFNN weight optimization have been recorded [13, 28, 29, 37]. The continuous version of the ant colony optimization (ACO) was efficiently applied to optimize the weight vectors of the FFNN by Socha and Blum [38]. Sarangi et al. [39] proposed the training of a feed-forward neural network using an artificial bee colony model with the back-propagation algorithm. In the study conducted by Kulluk et al. [40], the harmony search (HS) algorithm was employed to optimize the weight vector of the FFNNs. Ghasemiyeh et al. [45] presented a study that considered the prediction of stock exchange prices by proposing the implementation of a hybrid artificial neural network model with several metaheuristic algorithms. Their study used the cuckoo search algorithm, improved cuckoo search with genetic algorithm, and particle swarm optimization. Other global optimization metaheuristic techniques that have been similarly used for the FFNNs weight vectors and training optimization include the firefly algorithm (FA) [41], cuckoo search (CS) algorithm [42], gravitational search optimization (GSO) algorithm [43], and bacterial foraging optimization (BFO) algorithm [44]. However, few instance of the application of the symbiotic organisms search (SOS) algorithm for the training of other variants of ANNs exist in the literature (see [12, 32, 33]).

In this paper, to solve the stock market price prediction problem, an efficient hybrid symbiotic organisms search feedforward neural network, which uses the SOS algorithm to optimize the training and testing process of FFNNs, is proposed; here referred to as the SOSFFNN model. More so, the specific objective of the current study is to develop a new general hybrid framework that combines the global optimization metaheuristic approaches of SOS, PSO, and GA with the FFNNs model for effective and efficient prediction of stock price indices. Furthermore, the study also presents a comparative performance study of four different forecasting models, which include the ARIMA model, the proposed SOSFFNN model, particle swarm optimization trained feedforward neural networks (PSOFFNN) model and genetic algorithm trained feedforward neural networks (GAFFNN) model, of which the last three are hybrids of well-known global optimization metaheuristic algorithms with FFNNs. These three metaheuristic algorithms are selected based on their track records in terms of efficient performances, specific evolutionary operator characteristics, interesting collaborative interaction mechanisms between individuals or particles, efficient control parameter tuning and handling capability, stagnation prevention methods, good intensification and diversification balancing methods. For a fair comparison, the four aforementioned forecasting models are implemented and evaluated in parallel, in order to demonstrate the superior performance of the new proposed SOSFFNN model. Evaluation of the prediction accuracy of the proposed forecasting techniques are undertaken through the computation of the root mean squared error (RMSE), the mean absolute percentage error (MAPE) and the mean absolution deviation (MAD). In reality, this evaluation represents how much financial value each model possesses, as such performance is measured by profitability, consistency and robustness [9].

In summary, the main technical contributions of this paper are as follows:
- Design and implementation of a new SOS trained FFNN model to forecast the open and closed stock prices of various stock markets.
- Incorporation of SOS, GA, and SOS algorithms into the FFNN forecasting model to optimally find weights and biases for FFNN enhanced performance.
- Performance analysis of SOSFFNN, PSOFFNN, GAFFNN and ARIMA forecasting models on five markets using statistical indexes such as RMSE, MAPE, and including other performance indicators of profitability, consistency and robustness.

The rest of the paper is organized as follows: Section 2 presents related work, which involves using FFNNs and other hybrid models to perform the tasks of stock market forecasting. The representative metaheuristic algorithms approaches and their hybridization implementation techniques are discussed in Section 3. Section 4 presents the experimental results using recent time-series datasets. Finally, the concluding remarks are given in Section 5.

## 2. Related work

Recently, the application of artificial intelligence and machine learning techniques such as ANNs and support vector machines (SVM) have been used for the forecasting of straits time indices (STI). Several publications focus on portfolio optimisation, which is the act of selecting appropriate stocks in which to invest your money given finite capital and a finite set of available stocks [10,11]. Such portfolio optimization is a focus in using machine learning techniques for optimising return on investment. However, the focus of this study is different.,. The current study deals specifically with the idea of optimizing both training and learning performance of the ANNs using global optimization metaheuristic algorithms, as earlier mentioned. In this section, existing related studies are reviewed briefly and discussed to show the ideas underpinning the proposed forecasting methods.

Multiple methods for the prediction of stock prices has been investigated [9], using the stock prices of five different companies, which were obtained from Yahoo Finance. The four different forecasting methods investigated were the ARIMA model, ANNs, Holt-Winters (a statistical forecasting method for seasonal time-series data) and the time-series linear model (TSLM). It was found that the Holt-Winters method produced the best overall forecasting accuracy [9]. In [12], the SOS algorithm was proposed for. The computational results from training FFNNs with SOS were compared to results obtained from similar training of FFNNs using other metaheuristic search algorithms, such as the culture search(CS), genetic algorithm (GA), particle swarm optimisation (PSO), mean-variance optimisation (MVO), gravitational search algorithm (GSA), and biogeography-based optimisation (BBO). The results showed that the SOS trained the FFNNs the best for the task at hand [12].

The study conducted in [13] used both the PSO and backpropagation algorithms to train FFNNs for time-series forecasting. There were four types of time-series data used; specifically, for sunspots (number of sunspots observed over a period of time), exchange-rate (the USD to INR exchange rate), earthquakes (seismogram readings over time) and airline usage (the number of airline passengers). The results obtained were compared to results from other methods used to predict time-series data, such as the PSO-only trained FFNNs, backpropagation trained ANNs, and the Box-Jenkins models (which are statistically based models for predicting time-series data). Experimentation from the study showed that the PSO-only model was notably better than the backpropagation-only model and that the hybrid approach (PSO and backpropagation) was better than the Box-Jenkins models [13].

The FFNNs variants proposed in [14] were used for stock market data (NAV of SBI mutual fund) prediction and evaluation of the performances of three different methods for adjusting the network weights during training: the resilient backpropagation method, the Levenberg-Marquardt (also referred to as Bayesian regularisation) method and the scaled conjugate gradient method. It was observed that the Bayesian regularisation method was the best at being able to generalise the given data compared to the other training methods [14]. The study in [15] used the PSO algorithm to optimise the weights of an artificial neural network, which was used to forecast the exchange rate of the straits times index (STI) time series data. The results obtained were very promising and interesting. Therefore, building on the identified gap from the above related literature, the current study tries to replicate the earlier proposal made in [15] where the PSO algorithm was utilised for training neural networks.

Pillay and Ezugwu [32] proposed the implementation of a hybrid FFNN-based stock price prediction model that combines the FFNN and SOS algorithms. In [32], the standard SOS algorithm was used to optimize the weight vectors of the FFNN based model. The numerical results of the hybrid model showed that the SOS algorithm has some attractive potential that could further be extended and used successfully to optimize both the training and learning rate of the FFNNs. Therefore, based on the application of the SOS to train FFNN and also its superior prediction performance accuracy that was recorded in [32], the current study considers the possible extension of the aforementioned proposal, with the main goal of building a more robust and efficient stock price predictive model. Part of the extension also includes developing, in parallel, three other hybrid prediction models and validating them on a wide range of time series datasets. In addition, a comparative performance study of the four models developed in this paper is presented alongside results from the literature. In the next section, two of the main algorithms that inspired the current work, namely SOS and PSO, are briefly discussed. Thereafter, the implementation of the four hybrid models, SOSFFNN, PSOFFNN, GAFFNN, and ARIMA models is explained.

## 3. Representative metaheuristic algorithms

The symbiotic organisms search algorithm is a new metaheuristic optimization algorithm that has attracted the attention of the research community because of its simplicity of implementation and success records [16, 17, 47, 48, 49, 50]. The SOS algorithm has been applied widely, such as in the parallel machine scheduling problem [18, 19], optimal allocation of blood products [20, 21] and traveling salesman problem [22, 23]. The algorithm simulates the symbiotic interactions within a paired organism relationship, which are used to search for the fittest organism [17]. In the process of seeking the optimal global solution, the SOS iteratively uses a population of candidate solutions as promising areas of the search space. In the initial ecosystem, a group of organisms is generated randomly for the search space. Each organism represents one candidate solution and is associated with a certain fitness value, which reflects the degree of adaptation to the desired objective. The generation of new solutions is governed by three phases: the mutualism phase, commensalism phase, and parasitism phase. The nature of the interaction defines the main principle of each phase. Each organism interacts with the other organism randomly through all phases. In the mutualism phase, interactions benefit both sides; in the commensalism phase they benefit one side and do not impact the other and in the parasitism phase, interactions benefit one side and actively harm the other. The process is repeated until termination criteria are met. The reader may refer to the work presented in [17] for an in-depth understanding of the fundamental design concept and computational representation of the three SOS global optimization search phases.

The SOS algorithm was chosen because of its successful implementation in related research [17]. The SOS algorithm is perceived as capable of yielding good results and performance when applied to stock price, and also has the advantage, among many others, of its operations requiring no specific control parameter. These advantages were factored into the decision to consider this algorithm for the training of FFNNs. Furthermore, the algorithm avoids the risk of compromised performance due to improper parameter tuning, because only the parameters needing to be set are the size of the population or ecosystem and the maximum number of evaluations. This contrasts with other algorithms, such as the genetic algorithm (GA) or mine blast algorithm (MBA), or differential evolution (DE), PSO, or cuckoo search (CS), which all require at least more than one specific algorithm control parameter to be tuned, in addition to these two set parameters. The SOS algorithm uses three interaction strategies, mutualism, commensalism, and parasitism, to gradually improve the candidate solutions. This makes the algorithm simpler and quicker to implement since no time needs to be spent on the choice of operators. An organism (candidate solution) in this algorithm is represented by a vector of size 2, where the values are the opening and closing stock prices for a company. This representation was chosen due to the efficient way it represents all the necessary data and its ability to manipulate the data to get the best solutions.

The PSO algorithm used is the global best PSO hybridized with a neural network. In this algorithm, the neighborhood of each particle is the entire swarm. A swarm consists of a collection of particles, where each particle is a candidate solution. The particles are then evolved, where each particle's position and velocity are changed according to its own experience and that of its neighbors. Each particle can communicate with every other particle, and each particle is attracted to the best particle found by any other particle in the swarm. Each particle is a point in an n-dimensional space and contains the set of all the weights in the neural network and the bias. The algorithm stops when the maximum number of iterations has been reached. The position of the $i^{th}$ particle is represented as $x_i = x_{i1}, x_{i2}, \ldots, x_{in}$ and these components of position represent the individual weights and bias. The velocity of the $i^{th}$ particle is represented as $v_i = v_{i1}, v_{i2}, \ldots, v_{in}$. There are no selection or evolutionary operators that are used. Instead, the algorithm uses a fitness function with updates of positions and velocities to find near optimal solutions.

The PSO algorithm has been chosen as a candidate competitive algorithm for the proposed SOS algorithm because it is a common algorithm used for stock price predictions. It is a good algorithm to compare with SOS since PSO implementation with neural networks has already produced notable results for stock price prediction. It does not use operators such as mutation and crossover, which makes it simpler and easier to implement. The search can be carried out by the speed of the particle. During the development of several generations, only the most optimal particle can transmit information onto the other particles, and the speed of the researching is very fast. The global best PSO model has been chosen since it converges faster than the I-best or the local best PSO models. This is due to the larger particle interconnectivity of the global best PSO model. However, the global best PSO can easily be trapped in local minima, so more focus has to be given to exploration rather than exploitation during training. This is done by changing the PSO parameters, such as higher values for the maximum velocity and inertia weight.

### 3.1. Symbiotic organisms search trained neural network

In order to improve the predicting performance of the FFNN, it is hybridized with a SOS algorithm. The idea of hybridizing the SOS with a FFNN was motivated by similar implementation method presented in [15], in which the PSO was hybridized with a neural network. Therefore, since hybridizing PSO with a neural network seemed to be commonly reported in the literature as improving performance of the PSO, it sparked interest on how FFNN forecasting capability would perform if it was also hybridized with a neural network. Furthermore, stock price prediction may involve many companies' stocks and neural networks have good scalability to large datasets and work well with high dimensions. Neural networks also have the ability to model non-linear complex relationships, which, as already mentioned, is appropriate in the complexity of real-world stock market prediction. Accordingly, the application of neural networks hybridized with SOS would most likely be beneficial. This hybridization works by using the SOS algorithm to train the neural network by finding the optimal weights and biases for the network. This is similar to the way in which the PSO algorithm was used to train FFNNs [15]. The FFNN training method with an SOS, starts by normalizing and reading the required input of stock datasets, that is after the network has been structured by setting up the desired number of neurons in each of the three layers; namely, input, hidden and output layers. Then the training process can commence. For comparison, a generalized graphical representation of the three hybrid models, including SOSFFNN, PSOFFNN, and GAFFNN architectural frameworks, is given in Figure 1. It is important to note that each one of the metaheuristic algorithms, SOS, PSO, and GA, is employed to train the FFNNs and so three sets of weights and biases are simultaneously determined by these algorithms. This has the effect of the overall error of the individual FFNN and improving its corresponding accuracy in training the network. However, for the proposed SOSFFNN model, the structure of the FFNN is configured to be fixed.

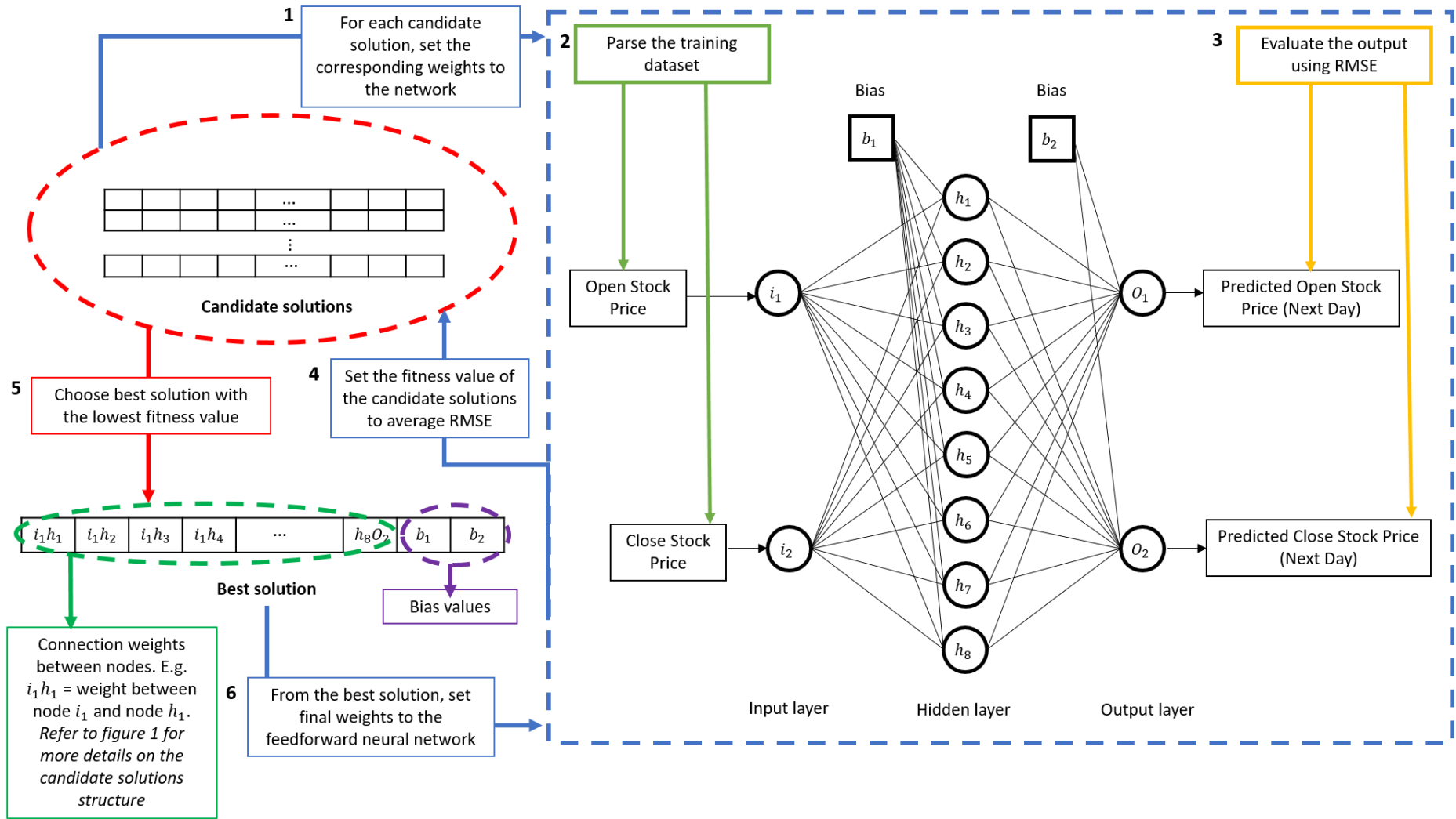

Fig. 1. Generalized architectural framework for FFNNs training by SOS, PSO, and GA metaheuristic algorithms

## 3.2. Solution representation

For the proposed SOSFFNN hybrid model, the neural network configuration is comprised of a single input layer with two nodes, a hidden layer with eight nodes and an output layer with two nodes. Each organism (candidate solution) is represented by a vector that contains the weights from the input layer to the hidden layer, the weights from the hidden layer to the output layer, and the bias value for the network. This vector has a length of 34. The representation of the vector is illustrated in Figure 2.

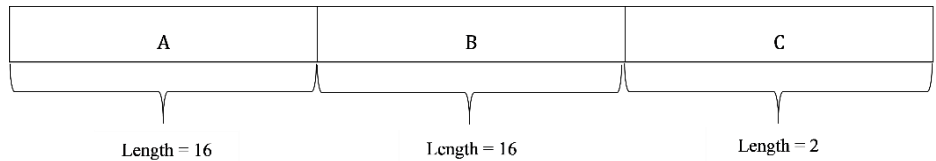

Fig. 2. Structure (Design) of Candidate Solution

A – weights linking the input layer nodes to the hidden layer nodes
B – weights linking the hidden layer nodes to the output layer nodes
C – the values for the bias

The algorithm is trained independently for each dataset comprising 1259 instances. The dataset is split as 80% for training and 20% for testing. After the train-test spilt, the data is normalized. A population size of 30 is used and the algorithm is run over 1000 iterations. The algorithm takes two inputs: the open price and the close price for a stock and it predicts these two prices for the next day. The RMSE is used as the fitness function in this algorithm; an error formula is judged as appropriate because the goal is to minimize the error of the prediction. The flowchart and pseudocode depicting the training process for the proposed hybrid SOSFFNN model are shown in Figure 3 and algorithm listing 1.

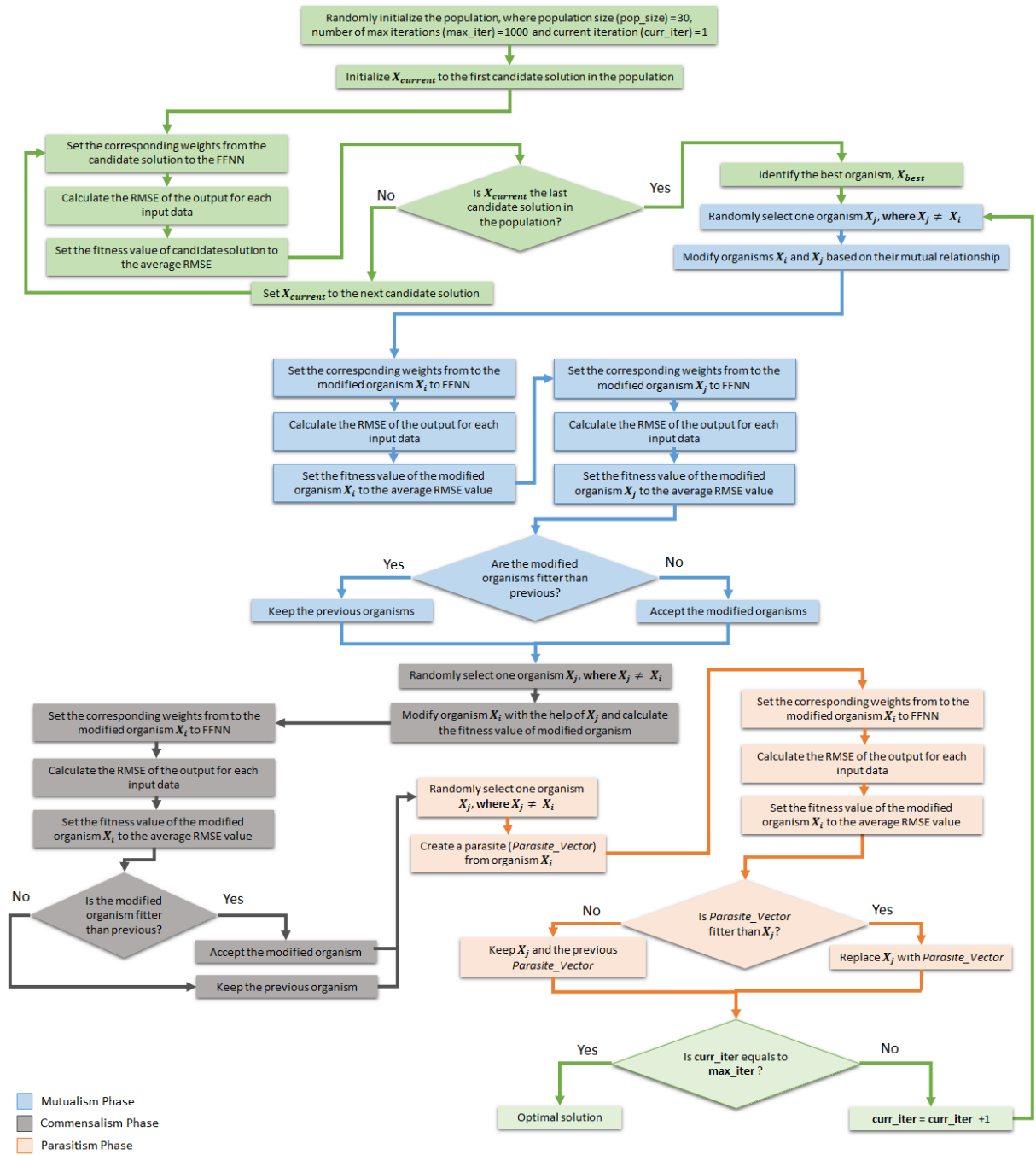

Fig. 3. Flowchart describing the process of training FFNN with SOS

**Algorithm 1:** Symbiotic organisms search trained feedforward neural network

| | |
|---|---|
| 1 | Initialize a population of size 30, composing of individuals described in Figure 1 (above). Each cell of the individuals is randomly initialized to values that are between 0 and 1. |
| 2 | REPEAT for each individual in the population |
| 3 |     Initialize the weights of the neural network with the corresponding weights contained by the individual |
| 4 |     REPEAT for each training instance |
| 5 |         Input the instance in the FFNN to obtain an output |
| 6 |         Calculate RMSE for the output and the expected output |
| 7 |     END |
| 8 |     Calculate the Average of RMSE values which will serve as the fitness value of the individual |
| 9 | END |
| 10 | REPEAT |
| 11 |     Increase number of iterations by 1 |
| 12 |     REPEAT for each individual $X_i$ in the population |
| 13 |         Set the best individual $X_{best}$ to the individual with the lowest fitness value |
| 14 |         MUTUALISM PHASE |
| 15 |             Select an individual $X_j$ randomly<br>Determine a mutual relationship vector<br>Mutual_Vector = $(X_i + X_j) / 2$ |
| 16 |             Determine the benefit factors BF1 and BF2, where the benefit factors are either 1 or 2 |
| 17 |             Modify $X_{i\_new}$ and $X_{j\_new}$ based on their mutual relationship<br>$X_{i\_new} = X_i + \text{rand}(0,1) * (X_{best} - \text{Mutual\_Vector} * BF1)$<br>$X_{j\_new} = X_j + \text{rand}(0,1) * (X_{best} - \text{Mutual\_Vector} * BF2)$ |
| 18 |             Calculate the fitness of $X_{i\_new}$ and $X_{j\_new}$ by using lines 3 to 8 |
| 19 |             IF $X_{i\_new}$ fitness value is less than $X_i$ |
| 20 |                 Replace $X_i$ with $X_{i\_new}$ |
| 21 |             END |
| 22 |             IF $X_{j\_new}$ fitness value is less than $X_j$ |
| 23 |                 Replace $X_j$ with $X_{j\_new}$ |
| 24 |             END |
| 25 |         END |
| 26 |         COMMENSALISM PHASE |
| 27 |             Select an individual $X_j$ randomly |
| 28 |             $X_{new} \leftarrow X_i + \text{rand}(-1,1) * (X_{best} - X_j)$ |
| 29 |             Calculate the fitness of $X_{new}$ |
| 30 |             IF $X_{i\_new}$ fitness value is less than $X_i$ |
| 31 |                 Replace $X_i$ with $X_{new}$ |
| 32 |             END |
| 33 |         END |
| 34 |         PARASITISM PHASE |
| 35 |             Select an individual $X_j$ randomly |
| 36 |             Create a parasite ($X_{parasite}$) from $X_i$ |
| 37 |             Calculate the fitness of $X_{parasite-+}$ |
| 38 |             IF $X_{parasite}$ fitness value is less than $X_j$ |
| 39 |                 Replace $X_j$ with $X_{parasite}$ |
| 40 |             END |
| 41 |         END |
| 42 |     END |
| 43 | UNTIL number of iterations are equal to 1000 |

### 3.3. Particle swarm optimization trained neural network

The second model implemented employs the PSO to train a FFNN. The neural network consists of an input layer with two nodes, a hidden layer comprising eight nodes and an output layer that has two nodes. The inputs are opening price for a stock and closing price for a stock for a day. The network outputs the predicted following day opening and closing prices for the stock. In this algorithm a swarm is initialized with 30 particles, where each particle is represented by a vector of size 34 that holds all the weights for the network as well as the bias value. The swarm is also initialized with random velocities. The $minX$ and $maxX$ values are set to -1 and 1, respectively. The fitness function used is RMSE, so that the error between the predicted values and the actual values can be minimized. The positions and velocities are updated for every iteration. The inertia value is set at 0.9, the two constants $c_1$ and $c_2$ are both set at 2 and the probability of death is 0.01. The algorithm runs to a maximum number of 1000 iterations. The best positions after the PSO is run provides the optimal weights for the neural network to be able to predict the output values. The flowchart and pseudocode for implementing PSOFFNN model are shown in Figure 4 and algorithm listing 2.

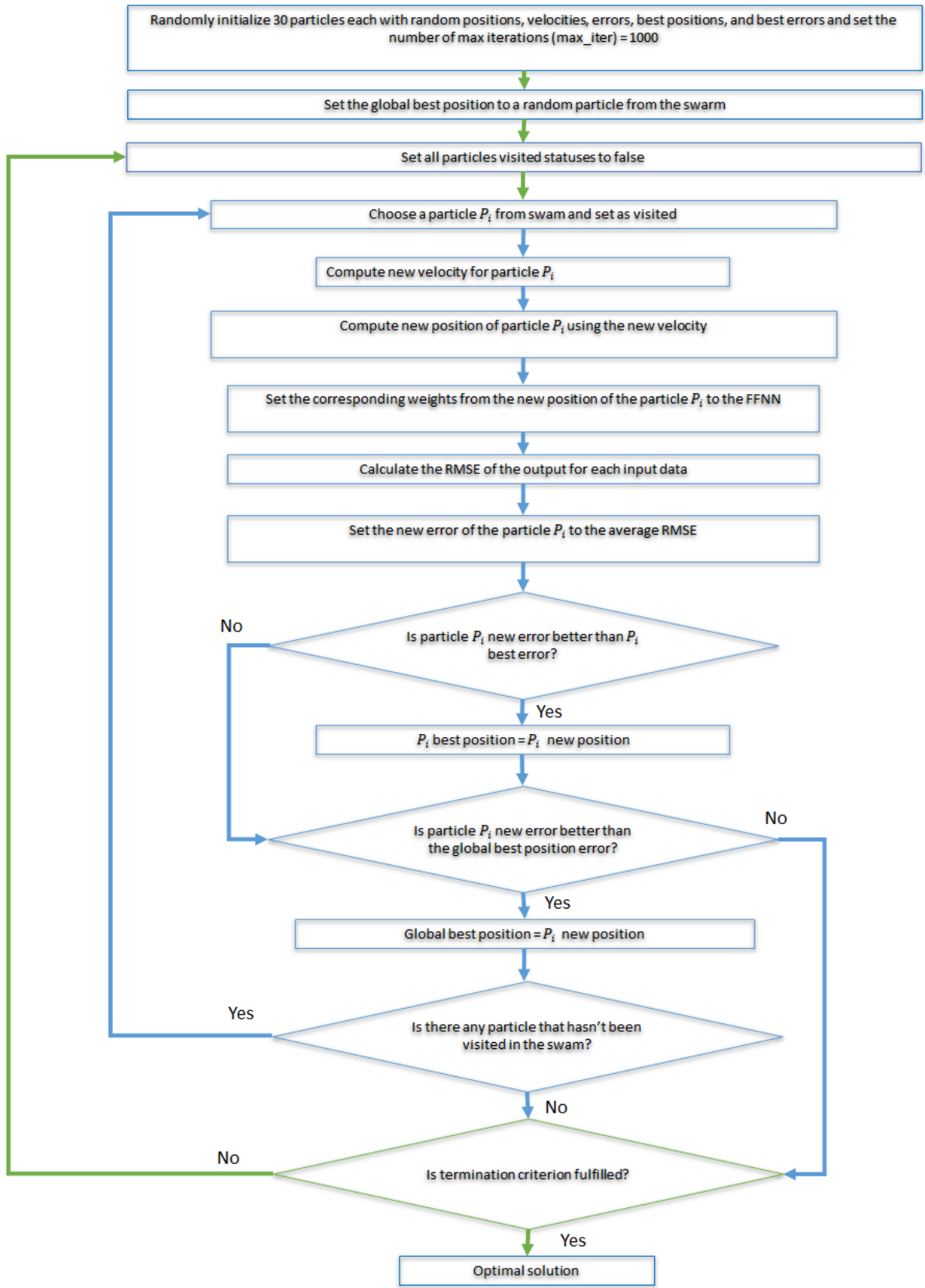

Fig. 4. Flowchart describing the process of training FFNN with PSO

**Algorithm 2:** Particle swarm optimization trained feedforward neural network

| | |
|---|---|
| 1 | Initialize number of particles to 30, and each particle $P_i$ to a random position, velocity, error, best position and best error |
| 2 | GlobalbestPosition ← Random particle from the swarm (population) |
| 3 | |
| 4 | REPEAT |
| 5 |     FOR EACH particle $P_i$ in swam |
| 6 |         Compute new velocity for particle $P_i$ |
| 7 |         Compute new position of particle $P_i$ using the new velocity |
| 8 |         Set the corresponding weights of the neural network using the new position of $P_i$ |
| 9 |         *Ev*aluate the neural network and compute the average RMSE of both the open and close predicted values |
| 10 | |
| 11 |         $P_i$ *new error* ← average RMSE of the neural network |
| 12 | |
| 13 |         IF particle $P_i$ new error < $P_i$ best error |
| 14 |             $P_i$ best position ← $P_i$ new position |
| 15 |         END |
| 16 | |
| 17 |         IF particle $P_i$ new error < $GlobalbestPosition$ error |
| 18 |             GlobalbestPosition ← $P_i$ new position |
| 19 |         END |
| 20 |     END |
| 21 | UNTIL number of iterations are equal to 1000 |

### 3.4. Genetic algorithm trained feedforward neural network

The third model implemented is the hybridization of GA with a feedforward neural network (GAFFNN). The operators used includes the roulette selection, uniform crossover with a crossover rate set to 0.5, and uniform mutation with a mutation rate set to $1/$(Length of individual) = 1/34. The population consisted of 30 individuals that were randomly initialized, and each individual is a vector of size 34. Elitism was used to select the best individual of the population for the next generation. The fitness function used for this algorithm is the RMSE because we aim to minimize the error between the actual and predicted values. This algorithm was allowed to run for a maximum of 1000 iterations, and the best individual served as the optimal weights for the neural network to predict the output values. The flowchart and pseudocode for implementing the hybrid GAFFNN model are shown in Figure 5 and algorithm listing 3.

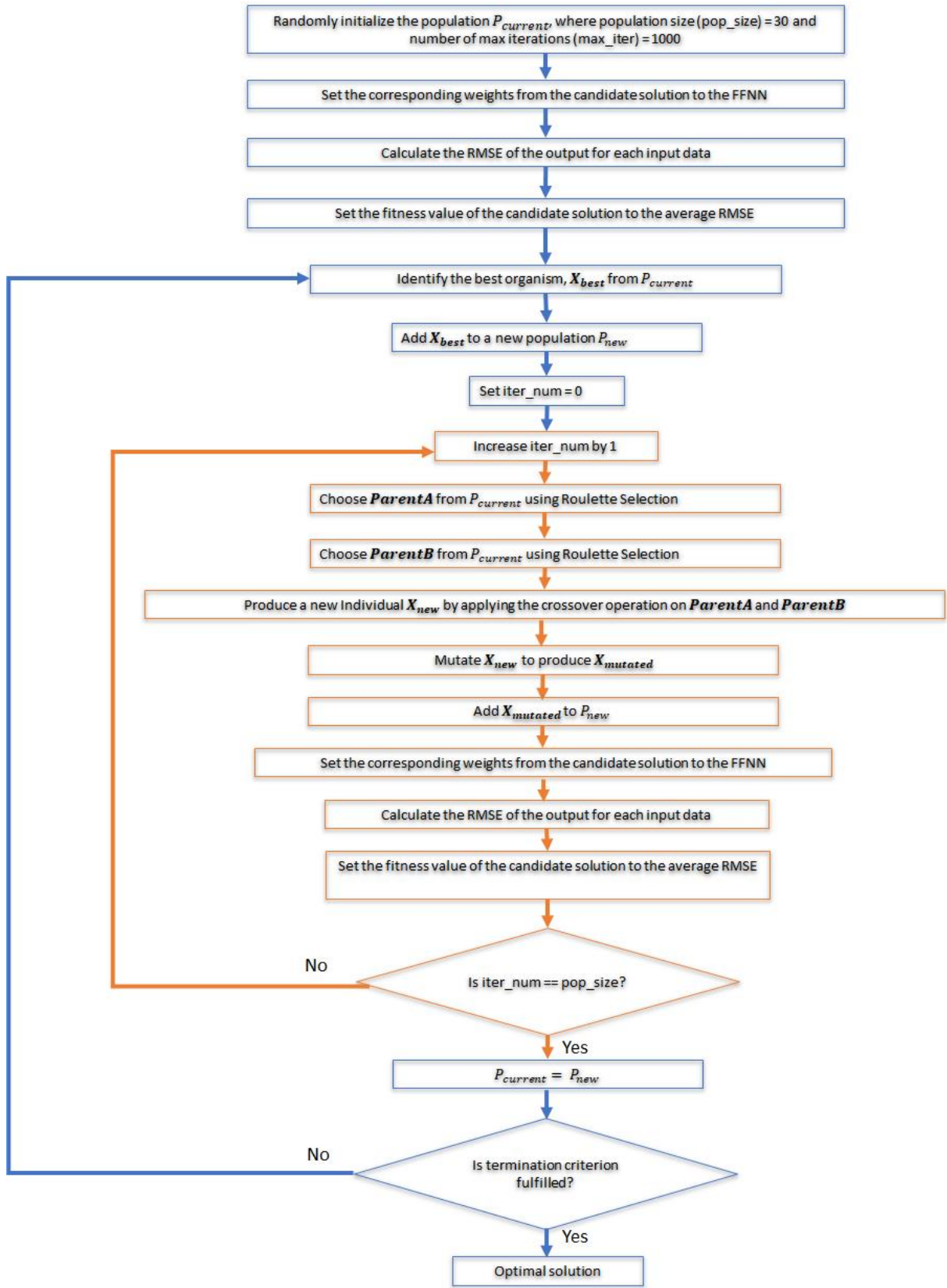

Fig. 5. Flowchart describing the process of training FFNN with GA

**Algorithm 3:** Genetic algorithm trained feedforward neural network

| | |
|---|---|
| 1 | **Initialize a population $P_{current}$ of size 30, composing of individuals described in Figure 1 (above).** |
| | Each cell of the individuals is randomly initialized to values that are between 0 and 1. |
| 2 | REPEAT for each individual in the population $P_{current}$ |
| 3 |     Initialize the weights of the neural network with the corresponding weights contained by the individual |
| 4 |     REPEAT for each training instance |
| 5 |         Input the instance in the FFNN to obtain an output |
| 6 |         Calculate RMSE for the output and the expected output |
| 7 |     END |
| 8 |     Calculate the Average of RMSE values which will serve as the fitness value of the individual |
| 9 | END |
| 10 | |
| 11 | REPEAT |
| 12 |     Increase number of iterations by 1 |
| 13 |     BestIndivdual ← The individual with the lowest fitness value from population $P_{current}$ |
| 14 |     Add BestIndivdual to a new population $P_{new}$ |
| 15 | |
| 16 |     REPEAT for size of the population $P_{current}$ |
| 17 |         ParentA ← Choose an individual from $P_{current}$ using Roulette Selection |
| 18 |         ParentB ← Choose an individual from $P_{current}$ using Roulette Selection |
| 19 | |
| 20 |         NewIndividual ← Perform crossover using ParentA and ParentB |
| 21 |         MutatedIndividual ← Perform mutation on NewIndividual |
| 22 | |
| 23 |         Calculate the fitness of the MutatedIndividual (using lines 3 to 9) |
| 24 |         Add MutatedIndividual to a new population $P_{new}$ |
| 25 | |
| 26 |     END |
| 27 |     $P_{current}$ ← $P_{new}$ |
| 28 | |
| 29 | UNTIL number of iterations are equal to 1000 |
| 30 | |
| 31 | BestIndivdual ← The individual with the lowest fitness value from population $P_{current}$ |
| 32 | Set the weights of the neural network with the corresponding weights contained by the BestIndividual |

### 3.5. Auto regressive integrated moving average

The fourth model is an autoregressive integrated moving average (ARIMA) model. This implementation was done with the extreme optimization numerical libraries for .NET. This library was built to assist developers to program financial, engineering and scientific applications. The auto regressive order was set to 0, the degree of differencing was set to 0 and the moving average was set to 2. These parameters yielded the best result compared to other combination of parameters settings tested on these datasets. Two ARIMA models were used; one to forecast each day's opening stock prices and the other the corresponding closing prices.

### 3.6. Evaluation metrics

A testing strategy that is used is the mean absolute percentage error (MAPE), which, in statistics, is a measure of the prediction accuracy of a forecasting method, for example in trend estimation. It is a very common testing strategy for stock price prediction algorithms and many organizations focus primarily on MAPE when assessing forecast accuracy. Most people are also more comfortable when dealing with percentage terms, which makes this error easy to interpret. The formula is GIVEN as follows:

$$MAPE = \frac{100\%}{n} \sum_{t=1}^{n} \left| \frac{A_t - F_t}{A_t} \right|,$$

(1)

where $A_t$ is the actual value of the stock price and $F_t$ is the forecast value from the algorithm. The absolute value in this calculation is summed for every forecasted point in time and divided by the number of fitted points, $n$. Multiplying by 100% makes it a percentage error. A drawback of this method is that it cannot be used for data that has zero values, since this could result in an error due to division by zero. This model is used nonetheless because it is highly unlikely that the price of a stock will be zero. Due to the pitfalls in MAPE, it is used in conjunction with other evaluation techniques like MAD (see below). With MAPE, the lower the percentage error the better.

Another common evaluation metric to test forecasting accuracy is the root mean squared error (RMSE). The RMSE is frequently used as a measure of the differences between values predicted by a model or an estimator and the values observed. This technique is used mainly when there is variance in the data, and it makes use of standard deviation. RMSE is the square root of the average of squared differences between a prediction and the actual observation. It expresses the average model prediction error in units of the variable of interest. The values from the metric can range from 0 to infinity and the direction of error is unspecified. The formula for RMSE is as follows:

$$RMSE = \sqrt{\frac{1}{n} \sum_{j=1}^{n} (y_j - \hat{y}_j)^2},$$

(2)

where $n$ is the number of values, $y_j$ is the forecast and the variable $\hat{y}_j$ is the mean error. Since the errors are squared before they are averaged, the RMSE gives a relatively high weight to large errors. This means that the RMSE would be more useful when large errors are particularly undesirable. RMSE avoids taking the absolute value. It is a negatively-oriented score, which means lower values are better.

The last testing metric discussed is the mean absolute deviation (MAD). Other than MAPE, MAD is the most popular metric for evaluating forecast accuracy. The mean absolute deviation of a dataset is the average distance between each data point and the mean. This strategy measures variance, just like RMSE, but lacks the strong statistical relationship of RMSE. In addition, MAD has the advantage of being easier to understand by those who are not specialists in the field, due to the error having the same dimension as the forecast. The formula for calculating MAD is as follows:

$$MAD = \frac{1}{T} \sum_{t=1}^{T} |e_t - \hat{e}_t|,$$

(3)

where $T$ is the number of time periods, $e_t$ is the forecast error in period $t$ and the last term denoted by $\hat{e}_t$ is the mean error for period $t$. The metric MAD is used in conjunction with MAPE to help overcome the pitfalls of MAPE and give a better overall view of the results.

The three testing metrics given above are used to ensure differing evaluations of the error of the forecast, thus making it easier to determine which forecasting algorithm produces the best results. This combined evaluation technique allows for a better comparison between the algorithms, so that a more informed decision can be made.

## 4. Experimental results

All the algorithms were run on the same MSI computer to allow for better comparison of the results. The computer specifications are as follows: Processor: Intel® Core™ i7-7700HQ, CPU @ 2.80GHz, Installed Memory (RAM): 12.0 GB, Graphics Card (GPU): Nvidia GTX1050, System Type: 64-bit Operating System, x64-based processor Operating System: Windows 10 Home. Each algorithm was run 10 times and the average of the results was recorded. The model implementation for this study was coded in C# using Microsoft Visual Studio 2017 as the IDE. The program has a GUI interface, where each of the three hybrid algorithms and ARIMA model can be run, and the results displayed upon completion of the program run. All the algorithms used the same dataset to facilitate meaningful performance comparisons between the algorithms. Furthermore, the proposed algorithms were run on the same computer for 1000 iterations with a population size of 30 ecosize. The design details of the SOS, PSO, and GA with neural network model and ARIMA model were presented in section 3.1.

## 4.1 Experimental result analysis

A total of sixteen stock indices are used to validate the performance quality of all the four models implemented in this study. The first five datasets used for the training and testing evaluation analysis of each model include the Straits Times Index, Nikkei 225, NASDAQ Composite, S&P 500, and Dow Jones Industrial Average financial stock data. The remaining eleven stock indices (presented in section 4.2) are used to validate the superior performance of the proposed SOSFFNN model against similar implementation results available from the literature. Three hybrid models, i.e. GAFFNN, SOSFFNN, and PSOFFNN, were executed 20 times for each dataset. A variety of combinations of parameters was investigated for the ARIMA model. The combination of parameters that achieved the best results across all datasets was used to compare with the results of the hybrid models. The parameters of the model are defined as follows: p – the lag order, d – the degree of differencing, q – the order of moving average. The results for each execution were evaluated using RMSE, MAPE and MAD. These evaluation metrics was applied to the opening and closing stock prices independently. Thereafter, the final RMSE, MAPE and MAD values was calculated by taking the average evaluations obtained for the opening and closing stock prices for each execution. Tables 1 to 5 below display the results obtained by the ARIMA model using different parameters on the above-mentioned datasets.

Table 1: Results of ARIMA model with different parameters executed on the Straits Times Index (STI) dataset

| (p, d, q) | RMSE | MAPE | MAD |
|---|---|---|---|
| (1, 0, 0) | 0.51496 | 726.47592 | 0.47442 |
| (1, 0, 1) | 0.47060 | 679.52866 | 0.43000 |
| (2, 0, 0) | 0.47093 | 679.88486 | 0.43034 |
| **(0, 0, 1)** | **0.27339** | **449.00106** | **0.23612** |
| (0, 0, 2) | 0.27285 | 449.12424 | 0.23544 |
| (1, 1, 0) | 0.58764 | 799.56756 | 0.54576 |
| (0, 1, 1) | 0.58756 | 799.48406 | 0.54567 |
| (1, 1, 2) | 0.58765 | 799.58167 | 0.54577 |
| (2, 1, 0) | 0.58779 | 799.72292 | 0.54591 |
| (2, 1, 2) | 0.58763 | 799.55608 | 0.54574 |
| (2, 1, 1) | 0.58765 | 799.58166 | 0.54577 |

Table 2: Results of ARIMA model with different parameters executed on the Nikkei 225 dataset

| (p, d, q) | RMSE | MAPE | MAD |
|---|---|---|---|
| (1, 0, 0) | 0.257656952 | 2357.559386 | 0.222398109 |
| (1, 0, 1) | 0.223440401 | 2179.917297 | 0.187451798 |
| (2, 0, 0) | 0.369366712 | 2862.846678 | 0.32351028 |
| **(0, 0, 1)** | **0.199942531** | **1468.738496** | **0.163198376** |
| (0, 0, 2) | 0.199744769 | 1468.634883 | 0.163052049 |
| (1, 1, 0) | 0.328223844 | 2684.223466 | 0.28749196 |
| (0, 1, 1) | 0.328222336 | 2684.217399 | 0.287490387 |
| (1, 1, 2) | 0.328244527 | 2684.306579 | 0.287513567 |
| (2, 1, 0) | 0.32822739 | 2684.237797 | 0.28749565 |
| (2, 1, 2) | 0.327707393 | 2682.18048 | 0.286949619 |
| (2, 1, 1) | 0.328243377 | 2684.302004 | 0.287512356 |

Table 3: Results of ARIMA model with different parameters executed on the NASDAQ Composite dataset

|         | RMSE      | MAPE         | MAD         |
|---------|-----------|--------------|-------------|
| (1, 0, 0) | 0.255949 | 769.0082205 | 0.194733137 |
| (1, 0, 1) | **0.235651** | 724.0124868 | **0.181368113** |
| (2, 0, 0) | 0.452104 | 1059.278054 | 0.374320556 |
| (0, 0, 1) | 0.36518  | 368.9350407 | 0.32365794  |
| (0, 0, 2) | 0.365181 | **368.9269246** | 0.323462226 |
| (1, 1, 0) | 0.258114 | 772.5081537 | 0.196327618 |
| (0, 1, 1) | 0.258459 | 773.0649823 | 0.196584668 |
| (1, 1, 2) | 0.259033 | 773.993644  | 0.197013238 |
| (2, 1, 0) | 0.259202 | 774.2607697 | 0.197138868 |
| (2, 1, 2) | 0.258456 | 773.0491356 | 0.196583485 |
| (2, 1, 1) | 0.259133 | 774.1481284 | 0.197087301 |

Table 4: Results of ARIMA model with different parameters executed on the S&P 500 dataset

| (p, d, q) | RMSE        | MAPE        | MAD         |
|-----------|-------------|-------------|-------------|
| (1, 0, 0) | 0.20447068  | 1669.778185 | 0.156086969 |
| (1, 0, 1) | 0.442889447 | **652.305845** | 0.401816475 |
| (2, 0, 0) | 0.259992093 | 1941.203308 | 0.196922849 |
| (0, 0, 1) | 0.390232405 | 801.2790239 | 0.355453868 |
| (0, 0, 2) | 0.3901787   | 801.2263174 | 0.35537639  |
| (1, 1, 0) | 0.203828916 | 1666.116668 | 0.155597012 |
| (0, 1, 1) | **0.203713539** | 1665.432994 | **0.155510929** |
| (1, 1, 2) | 0.204357768 | 1669.29952  | 0.155991186 |
| (2, 1, 0) | 0.204953521 | 1672.773333 | 0.156434643 |
| (2, 1, 2) | 0.204083968 | 1667.673622 | 0.155787515 |
| (2, 1, 1) | 0.204362647 | 1669.323426 | 0.155994957 |

Table 5: Results of ARIMA model with different parameters executed on the Dow Jones Industrial Average dataset

| (p, d, q) | RMSE      | MAPE         | MAD         |
|-----------|-----------|--------------|-------------|
| (1, 0, 0) | 0.199051  | 4912.181919  | 0.154297985 |
| (1, 0, 1) | **0.190526** | 4531.451637 | **0.153318065** |
| (2, 0, 0) | 0.291742  | 6146.048597  | 0.239788249 |
| (0, 0, 1) | 0.394956  | 2107.878253  | 0.361880174 |
| (0, 0, 2) | 0.394909  | **2107.870186** | 0.361829727 |
| (1, 1, 0) | 0.200285  | 4947.555069  | 0.154654101 |
| (0, 1, 1) | 0.200212  | 4946.549816  | 0.154600909 |
| (1, 1, 2) | 0.200929  | 4956.618305  | 0.155127791 |
| (2, 1, 0) | 0.201839  | 4969.060447  | 0.155820802 |
| (2, 1, 2) | 0.200679  | 4953.147935  | 0.154942359 |
| (2, 1, 1) | 0.200821  | 4955.118616  | 0.155047078 |

The ARIMA models with parameters p = 1, d = 0, q =1 produced good results for all datasets. Table 6 below presents, using the STI dataset, the best result, average and standard deviation from the hybrid models, and the corresponding results achieved by the ARIMA model. The SOSFFNN model obtained the lowest RMSE, MAPE and MAD values (the best results are highlighted in bold). However, all algorithms achieved RMSE and MAD value very close to zero, indicating very small prediction error. Judging by the average MAPE values, the SOSFFNN was the only method that received a MAPE value that is below 100%.

Table 6: Results obtained by the various models executed on the Straits Times Index (STI) dataset

|  | *RMSE* | | | |
| --- | --- | --- | --- | --- |
|  | **PSOFFNN** | **GAFFNN** | **SOSFFNN** | **ARIMA** |
| Best result | 0.098514582 | 0.08554277 | 0.056947606 | 0.27339 |
| Average | 0.153291915 | 0.15695206 | **0.068066492** | 0.27339 |
| Standard Deviation | 0.036538728 | 0.03731014 | 0.009085544 | - |
|  | **MAPE** | | | |
|  | **PSOFFNN** | **GAFFNN** | **SOSFFNN** | **ARIMA** |
| Best result | 140.7034154 | 68.6739057 | 61.00968446 | 449.00106 |
| Average | 181.5528373 | 162.905379 | **79.4198888** | 449.00106 |
| Standard Deviation | 90.60608094 | 54.4678486 | 31.70291803 | - |
|  | **MAD** | | | |
|  | **PSOFFNN** | **GAFFNN** | **SOSFFNN** | **ARIMA** |
| Best result | 0.073634308 | 0.06893376 | 0.045293643 | 0.23612 |
| Average | 0.131810335 | 0.14661241 | **0.054640045** | 0.23612 |
| Standard Deviation | 0.034575922 | 0.03762511 | 0.006675286 | - |

The SOSFFNN model also obtained the lowest RMSE, MAPE and MAD values on the Nikkei 225 dataset, as seen in Table 7. The MAPE values for all algorithms are well above 100%, but for the hybrid SOS algorithm an average MAPE was achieved that is much lower than that for the other three models. The MAPE values being above 100% simply means that the errors obtained are much greater than the actual values.

Table 7: Results obtained by the various models executed on the Nikkei 225 dataset

|  | *RMSE* | | | |
| --- | --- | --- | --- | --- |
|  | **PSOFFNN** | **GAFFNN** | **SOSFFNN** | **ARIMA** |
| Best result | 0.062963847 | 0.069220672 | 0.048155144 | 0.223440401 |
| Average | 0.120360658 | 0.126404491 | **0.0639251** | 0.223440401 |
| Standard Deviation | 0.027234249 | 0.027360186 | 0.01055843 | - |
|  | **MAPE** | | | |
|  | **PSOFFNN** | **GAFFNN** | **SOSFFNN** | **ARIMA** |
| Best result | 536.1621864 | 565.1700435 | 344.4495285 | 2179.917297 |
| Average | 737.255461 | 865.5378591 | **382.960241** | 2179.917297 |
| Standard Deviation | 395.5116949 | 224.7175003 | 104.242976 | - |
|  | **MAD** | | | |
|  | **PSOFFNN** | **GAFFNN** | **SOSFFNN** | **ARIMA** |
| Best result | 0.047500211 | 0.053351207 | 0.036292922 | 0.187451798 |
| Average | 0.099211153 | 0.110262775 | **0.048235759** | 0.187451798 |
| Standard Deviation | 0.025993205 | 0.027168657 | 0.00911356 | - |

Table 8: Results obtained by the various models executed on the S&P 500 dataset

|  | RMSE | | | |
| --- | --- | --- | --- | --- |
|  | **PSOFFNN** | **GAFFNN** | **SOSFFNN** | **ARIMA** |
| Best result | 0.07628775 | 0.0931389 | 0.0787657 | 0.442889447 |
| Average | 0.19994111 | 0.1142055 | **0.0981264** | 0.442889447 |
| Standard Deviation | 0.06892206 | 0.0156173 | 0.0114496 | - |
|  | MAPE | | | |
|  | **PSOFFNN** | **GAFFNN** | **SOSFFNN** | **ARIMA** |
| Best result | 39.9486099 | 650.79688 | 369.64911 | 652.305845 |
| Average | 438.293708 | 677.07549 | **338.09731** | 652.305845 |
| Standard Deviation | 264.710215 | 141.94039 | 94.296426 | - |
|  | MAD | | | |
|  | **PSOFFNN** | **GAFFNN** | **SOSFFNN** | **ARIMA** |
| Best result | 0.04782616 | 0.0753791 | 0.0746563 | 0.401816475 |
| Average | 0.17832244 | 0.0978768 | **0.0865577** | 0.401816475 |
| Standard Deviation | 0.10639377 | 0.0185092 | 0.0109373 | - |

The hybridized PSOFFNN model best result on the S&P 500 dataset were in each case the lowest amongst the models, which can be seen above in Table 8. The PSOFFNN best result received a MAPE value that was well below 100%, whist the other models obtained MAPE values well above 100%. However, the SOS hybrid model achieved the best average results for RMSE, MAPE and MAD on this dataset. This result concerning the PSOFFNN hybrid model assures us that PSO can also achieve remarkable results in making better parameter tuning, i.e. population size and number of max iterations. The RMSE and MAD values for all algorithms are close to zero, with the hybrid models obtaining values very much closer to zero than those for the ARIMA model.

Table 9: Results obtained by the various models executed on the NASDAQ Composite dataset

|  | RMSE | | | |
| --- | --- | --- | --- | --- |
|  | **PSOFFNN** | **GAFFNN** | **SOSFFNN** | **ARIMA** |
| Best result | 0.084575 | 0.081267 | 0.0847377 | 0.235651 |
| Average | 0.1700351 | 0.115943 | **0.1044186** | 0.235651 |
| Standard Deviation | 0.08755651 | 0.01794 | 0.0388474 | - |
|  | MAPE | | | |
|  | **PSOFFNN** | **GAFFNN** | **SOSFFNN** | **ARIMA** |
| Best result | 214.994576 | 124.2538 | 97.130579 | 724.0124868 |
| Average | 182.493572 | 175.3707 | **76.878466** | 724.0124868 |
| Standard Deviation | 65.3674547 | 59.31463 | 23.405629 | - |
|  | MAD | | | |
|  | **PSOFFNN** | **GAFFNN** | **SOSFFNN** | **ARIMA** |
| Best result | 0.06681631 | 0.099005 | 0.0694186 | 0.181368113 |
| Average | 0.14740783 | 0.097637 | **0.0859739** | 0.181368113 |
| Standard Deviation | 0.08329605 | 0.013304 | 0.0348924 | - |

As can be seen in Table 9, the PSOFFNN hybrid model, yet again, was able to achieve the best result; that is the lowest RMSE and MAD values for the NASDAQ Composite dataset. However, these values were slightly lower than those values of the hybridized SOSFFNN algorithm. The hybrid SOS model best result for MAPE was, nevertheless, the lowest amongst the remaining models; being the only MAPE value below 100%. Ultimately, the SOSFFNN hybrid model achieved the lowest average RMSE, MAPE and MAD results.

Table 10, below, shows the results obtained by the different models on the Dow Jones Industrial Average dataset. The hybrid SOSFFNN model achieved the best results. All algorithms produced MAPE values well above 100%,

but the SOS hybrid model produced the lowest compare to the models. The RMSE and MAD values are close to zero which indicates the algorithms' low prediction errors.

Table 10: Results obtained by the various models executed on the Dow Jones Industrial Average dataset

|  | *RMSE* | | | |
| --- | --- | --- | --- | --- |
|  | **PSOFFNN** | **GAFFNN** | **SOSFFNN** | **ARIMA** |
| Best result | 0.079243065 | 0.0874652 | 0.07653209 | 0.190526 |
| Average | 0.188904686 | 0.1152931 | **0.09272649** | 0.190526 |
| Standard Deviation | 0.087414082 | 0.0184702 | 0.01034236 | - |
|  | **MAPE** | | | |
|  | **PSOFFNN** | **GAFFNN** | **SOSFFNN** | **ARIMA** |
| Best result | 1528.434177 | 1812.1675 | 785.352258 | 4531.451637 |
| Average | 1333.015383 | 1909.0536 | **841.44458** | 4531.451637 |
| Standard Deviation | 564.8132856 | 397.32541 | 141.331854 | - |
|  | **MAD** | | | |
|  | **PSOFFNN** | **GAFFNN** | **SOSFFNN** | **ARIMA** |
| Best result | 0.065133517 | 0.0719295 | 0.05989382 | 0.153318065 |
| Average | 0.169147186 | 0.0957098 | **0.0739986** | 0.153318065 |
| Standard Deviation | 0.085906861 | 0.0159037 | 0.00908367 | - |

Figure 6 illustrates a sample system implementation test run for the developed stock price prediction simulator software. As mentioned earlier, the developed stock price prediction system has a graphical user interface that consists of a dropdown list, where users can select any one of the forecasting models. The simulator also has a feature that provides for dataset upload, whereby both the training and testing data samples can be uploaded to the system for processing. The entire processing task is achieved by iteratively executing each model for 10 replications over 1000 iterations, after which the processed results are generated and displayed as shown in Figure 6.

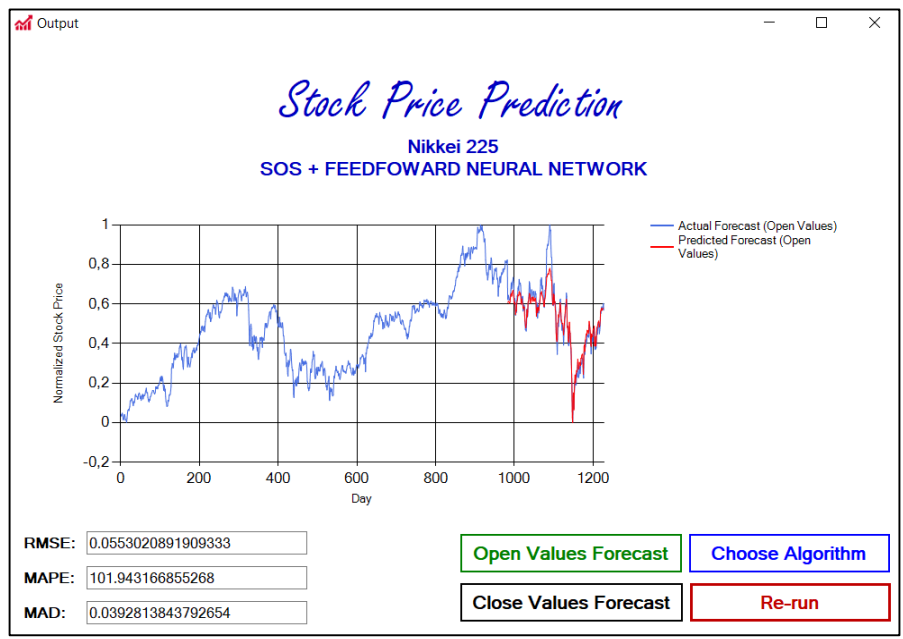

Fig. 6. Stock price prediction user interface with a graph displaying the open prices prediction forecast of the SOSFFNN hybrid model on the Nikkei 225 dataset

For each day, the forecast gives predictions for both opening and closing stock prices of the next day, as presented in Figures 7 to 26 for the different datasets. In the graphs, the y-axis depicts the normalized stock prices

(opening/closing), and the x-axis represents the different days in chronological order. Also each figure depicts the actual (denoted with a blue color) and predicted (denoted with a red color) values forecasts as a time series. It is important to note that in the case of an ideal prediction graph, the values are more inclined toward the zero margin. This is because any deviations from zero clearly indicates a deviation from a good prediction by the model [46]. Subsequently, the results of actual forecast versus predicted forecast for all the experimented datasets using the four models are analyzed.

The Singapore stock market uses the Straits Times Index (STI) as a benchmark to track the performance of the top 30 companies listed on its exchange. The ARIMA, GAFFNN, PSOFNN and the proposed SOSFFNN prediction models were executed on the STI dataset and its predictive performances capabilities are illustrated in Figures 7 to 10, respectively. The ARIMA model displays the worst performance when compared to the other models. It over predicts both the opening and closing stock prices for each day. Both predicted forecast graphs illustrated in Figure 7 show a slope with a negative gradient and with decreasing steepness. The shape of both predicted forecast graphs clearly does not correlate with the shape of the actual forecast graphs, which further indicates the ARIMA model's poor performance. Contrasting with this poor model performance, the fluctuations of GAFFNN, PSOFFNN and the SOSFFNN models' predicted forecast graphs are similar to the fluctuations of the actual forecast graphs. These fluctuations are important, because they serve as an indication to investors that there could possibly be a rise or fall in the stock price. Comparing the GAFFNN and PSOFFNN models' stock price predicted values, the GAFFNN performs better than the PSOFFNN model. The predicted stock price forecast results of the GAFFNN are much closer to the actual stock prices, as shown in Figure 8. In comparing the GAFFNN and SOSFFNN models, the GAFFNN model does produce a better prediction from Day 1000 to around Day 1030. Thereafter, the SOSFFNN performs much better than the GAFFNN. This can be seen by the predicted forecasts being positioned extremely close to the actual forecasts graphs. Hence, the SOSFFNN predictive performance dominates over the other models on the STI dataset.

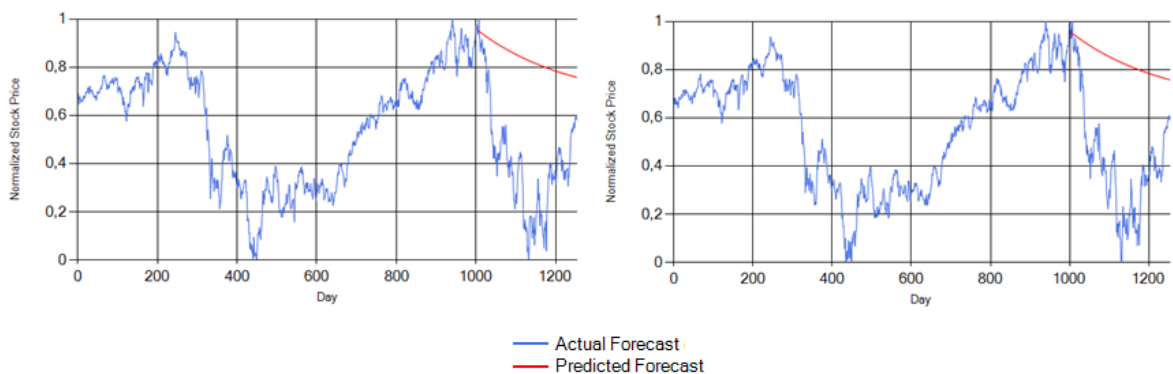

Fig. 7. Graphs displaying the open stock price (left graph) and close stock price (right graph) prediction forecast obtained by the ARIMA model on the Straits Times Index dataset

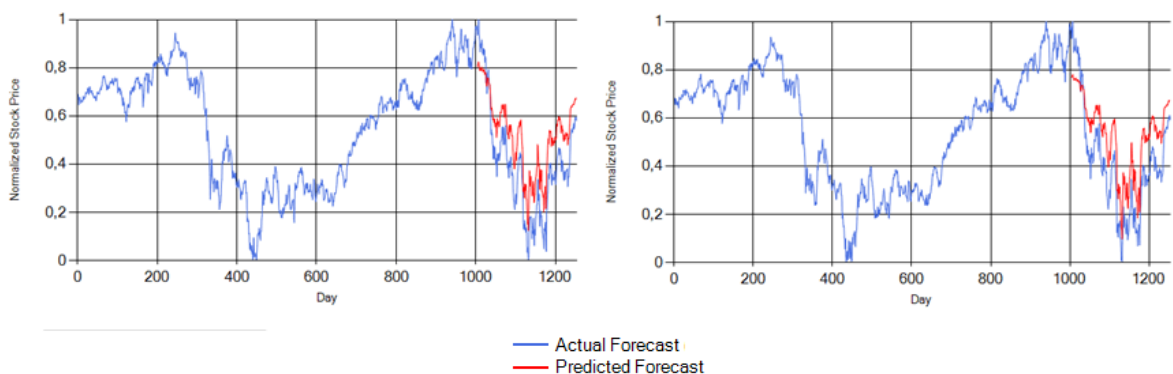

Fig. 8. Graphs displaying the open stock price (left graph) and close stock price (right graph) prediction forecast obtained by the GAFFNN model on the Straits Times Index dataset

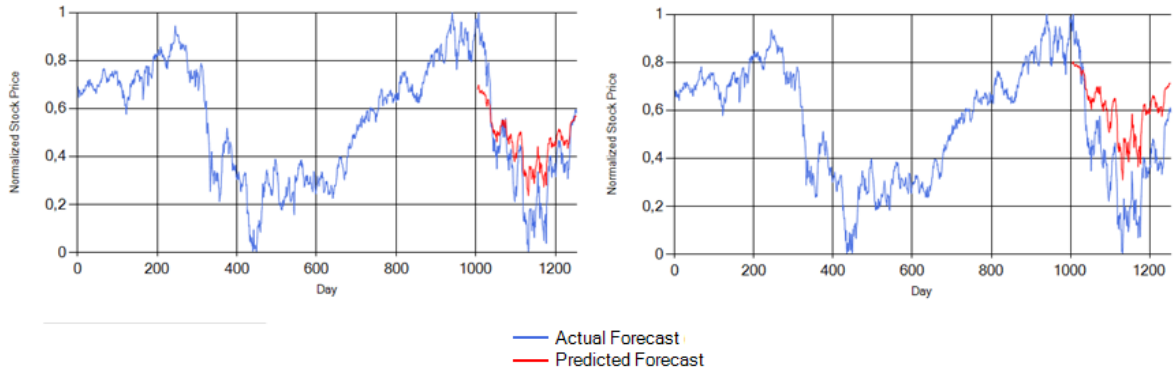

Fig. 9. Graphs displaying the open stock price (left graph) and close stock price (right graph) prediction forecast obtained by the PSOFFNN model on the Straits Times Index dataset

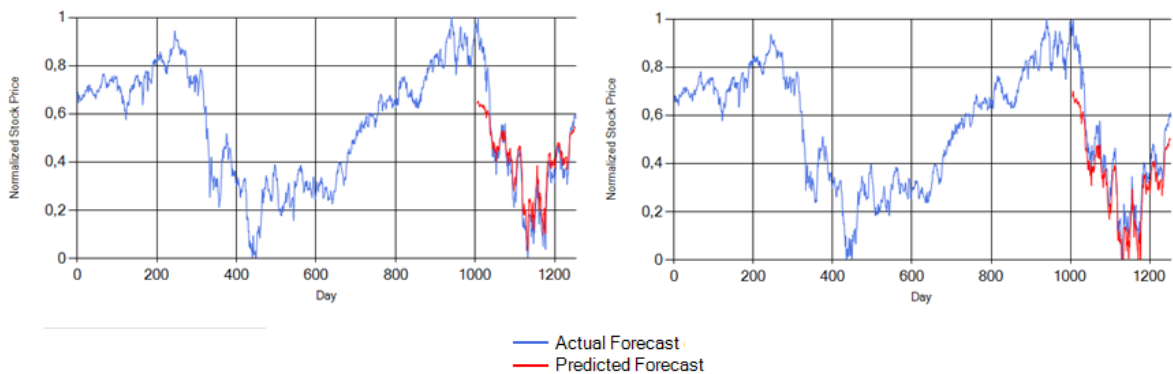

Fig. 10. Graphs displaying the open stock price (left graph) and close stock price (right graph) prediction forecast obtained by the SOSFFNN model on the Straits Times Index dataset

The Nikkei 225 is a stock market index for the Tokyo stock exchange. All the prediction models were executed on the Nikkei 225 dataset and the results are shown in Figures 11 to 14. The ARIMA model has, yet again, displayed the worst performance of all the models on this dataset. The predicted forecast graphs tend to fit the trend of the local peaks of the actual stock price forecasts. Even though this is the case, the predicted forecast only informs the investor of the probability that there will be a fall in the stock market price. By contrast, the other models display predicted forecast fluctuation trends that are similar to those of the actual stock price graphs. All the hybrid models show very similar prediction performances on this dataset. However, a critical evaluation of these forecasts shows that, around the $1150^{th}$ day, the predicted opening stock prices of the PSOFFNN are much closer to the actual stock prices than they are for the GAFFNN. However, the SOSFFNN illustrates a predicted opening stock price forecast that sits almost exactly on the actual opening stock price. Comparing the values forecast for closing stock price, up to the $1100^{th}$ day, the predicted stock prices of the PSOFFNN are much closer to the actual stock price than are those for the GAFFNN. However, after that day GAFFNN predicted stock prices are much closer to the actual stock prices. Nonetheless, the SOSFFNN illustrates a predicted closing stock price forecast that sit almost exactly on the actual opening stock price. From the results of this dataset, it can be gathered that GAFFNN performs better in some cases than the PSOFFNN and vice versa. This highlights the need for future work in observing the trend of results generated from the combination of both genetic algorithm and particle swarm optimization to train the feedforward network.

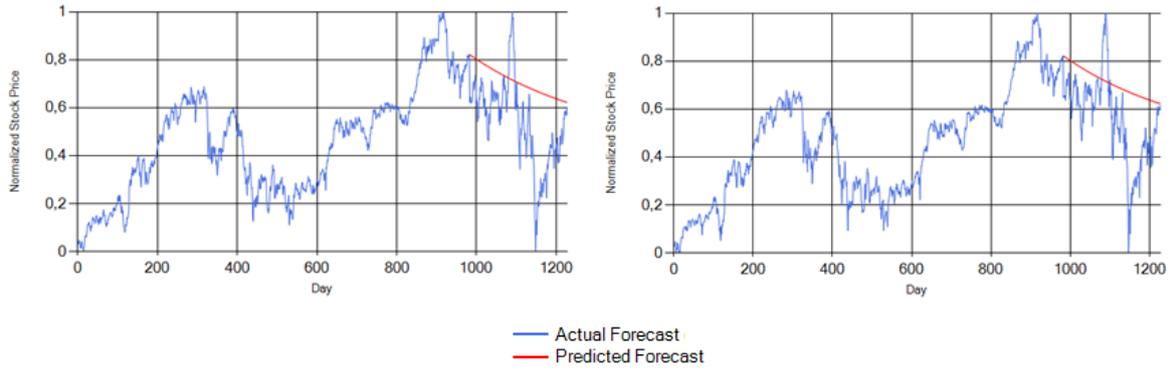

Fig. 11. Graphs displaying the open stock price (left graph) and close stock price (right graph) prediction forecast obtained by the ARIMA model on the Nikkei 225 dataset

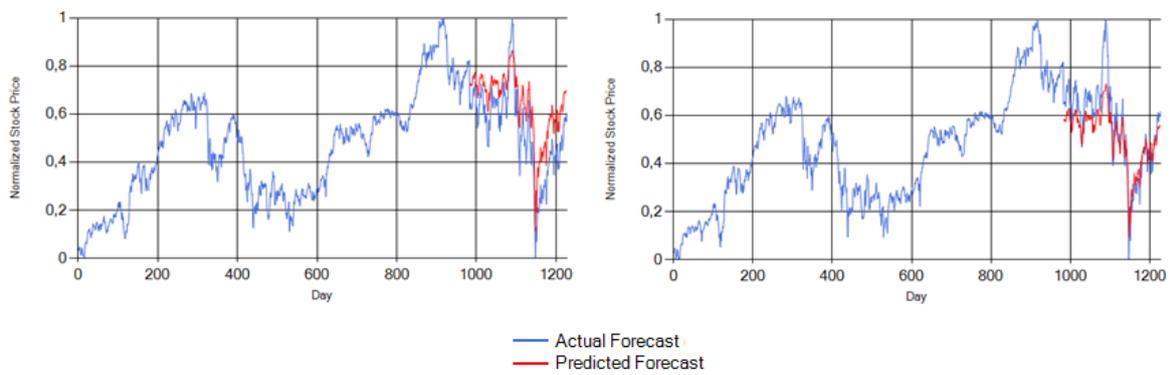

Fig. 12. Graphs displaying the open stock price (left graph) and close stock price (right graph) prediction forecast obtained by the GAFFNN model on the Nikkei 225 dataset

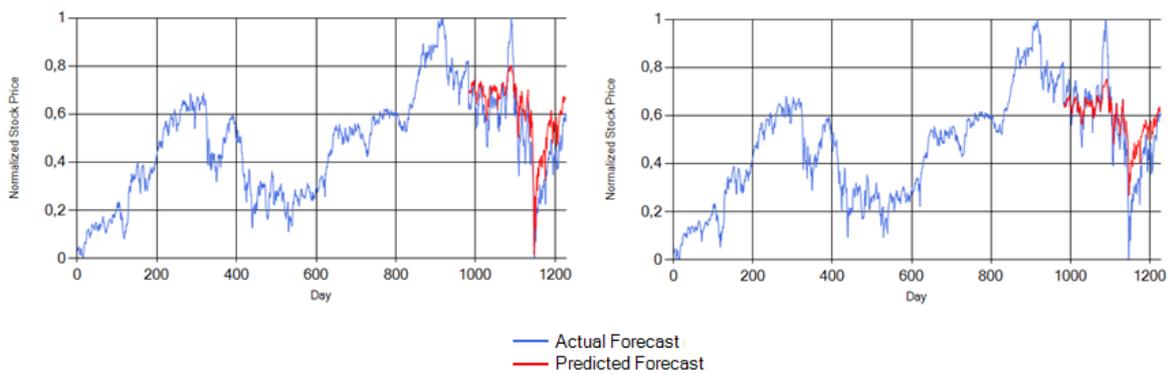

Fig. 13. Graphs displaying the open stock price (left graph) and close stock price (right graph) prediction forecast obtained by the PSOFFNN model on the Nikkei 225 dataset

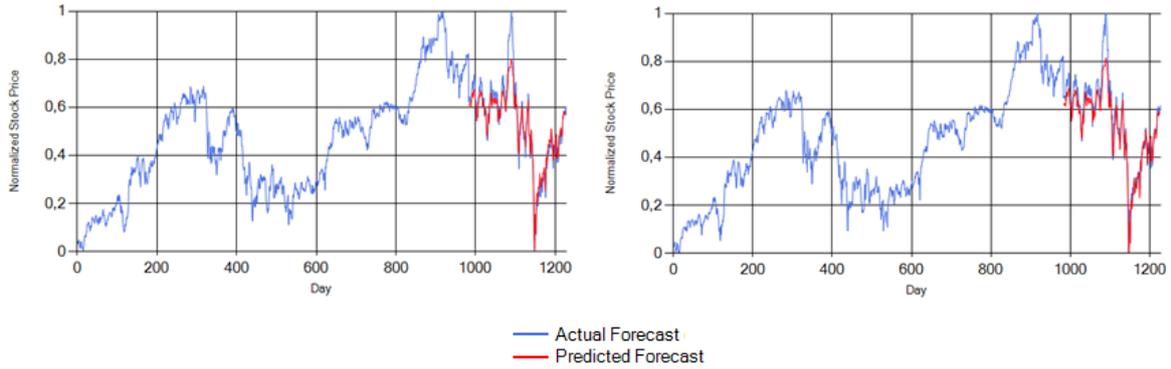

Fig. 14. Graphs displaying the open stock price (left graph) and close stock price (right graph) prediction forecast obtained by the SOSFFNN model on the Nikkei 225 dataset

The S&P500 index measures the stock performance of 500 companies list on the United States stock exchange. It is considered to be the best representation of the United States stock market. All the prediction models were executed on the S&P500 dataset and the forecasting results are presented in Figures 15 to 18 respectively. The ARIMA model's predicted forecast graphs take on a different shape from the other ARIMA model's forecasts graphs in earlier figures. The forecast graphs begin with a very steep slope and then the slope changes to being close to zero after day 1050. This shows that the majority of the stock prices are significantly lower than the actual stock price. On the other hand, the PSOFFNN model seems to experience some difficulty predicting both the opening and closing stock prices, because it over predicts the opening stock prices and under predicts the closing stock prices. While both the GAFFNN and SOSFFNN models show better prediction performances than for PSOFFNN, the SOSFFNN, nevertheless, illustrates a slightly better performance than does the GAFFNN. The SOSFFNN predicted stock prices are much closer to the actual stock prices, and also it shows a very similar trend of fluctuations compared to the GAFFNN model's predicted forecast.

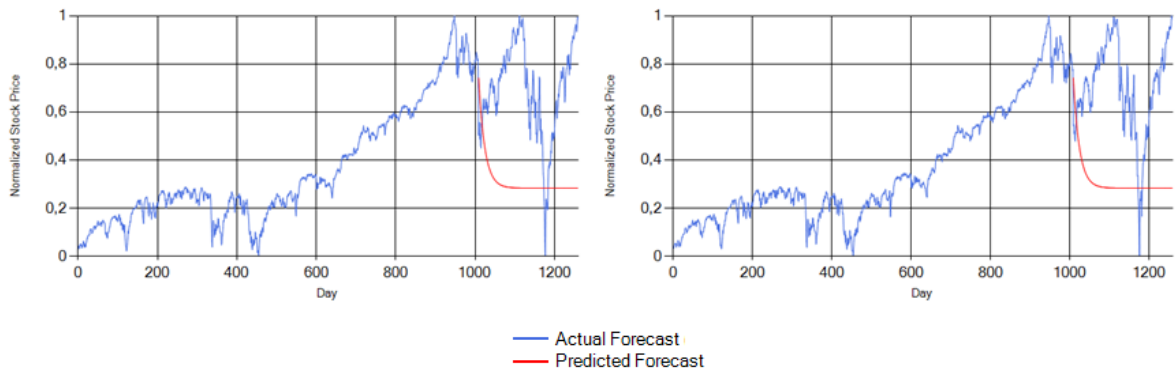

Fig. 15. Graphs displaying the open stock price (left graph) and close stock price (right graph) prediction forecast obtained by the ARIMA model on the S&P 500 dataset

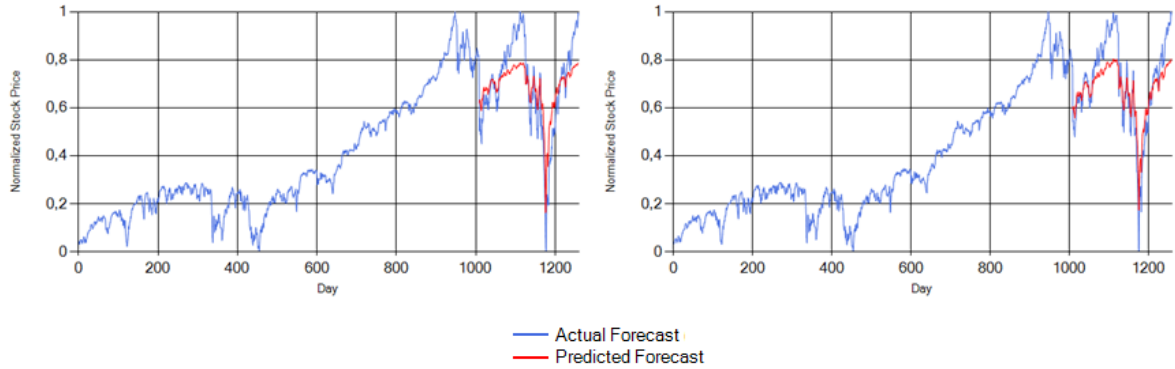

Fig. 16. Graphs displaying the open stock price (left graph) and close stock price (right graph) prediction forecast obtained by the GAFFNN model on the S&P 500 dataset

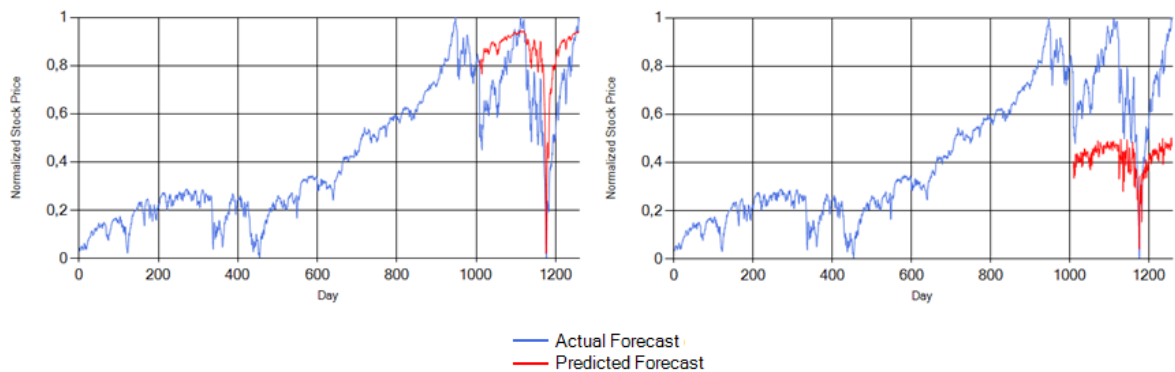

Fig. 17. Graphs displaying the open stock price (left graph) and close stock price (right graph) prediction forecast obtained by the PSOFFNN model on the S&P 500 dataset

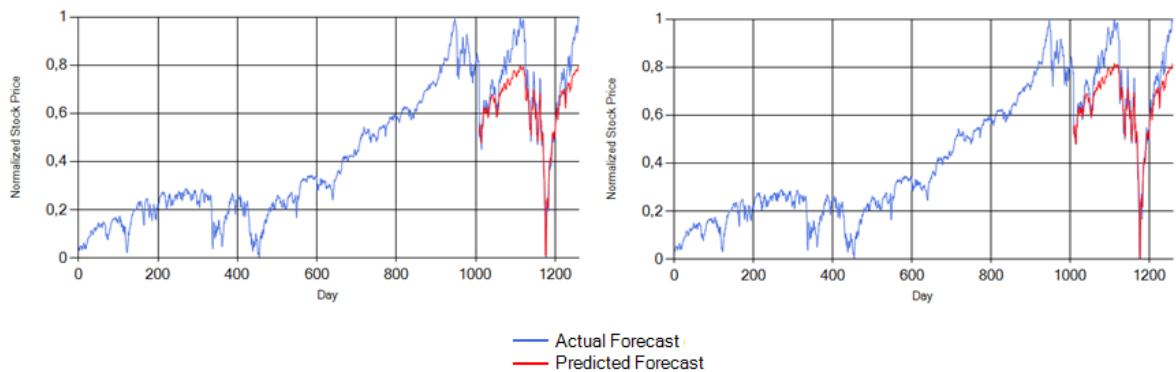

Fig. 18. Graphs displaying the open stock price (left graph) and close stock price (right graph) prediction forecast obtained by the SOSFFNN model on the S&P 500 dataset

The Dow Jones Industrial Average is a stock market index that measures the performances of the top 30 companies listed on the United States stock exchange. The ARIMA, GAFFNN, PSOFNN and the SOSFFNN models were executed on the Dow Jones Industrial Average dataset and the results are shown in Figures 19 to 22, respectively. Again, the ARIMA model has displayed the worst performance among the models. Both the predicted stock price forecast graphs in Figure 19 illustrate a curve that cuts through the actual stock price forecast graphs. The shape of these graphs does not correlate with the shape of the actual stock price forecast graphs, indicating the ARIMA model's poor performance. Since, the predicted stock price forecast graphs tend to have a constant negative slope, the overall prediction is of less important to investors. The fluctuations of GAFFNN, PSOFFNN and the

SOSFFNN models' predicted forecast graphs are similar to the fluctuations of the actual stock price forecast graphs. The opening stock price prediction forecasts tend to be similar across all three hybrid models. But, a clear distinction in the prediction performance emerges when comparing the predicted forecasts of closing stock prices. Overall, the GAFFNN and SOSFFNN models predict stock prices lower than the actual stock prices. Whilst the PSOFFNN predicts the majority of the stock price to be higher than the actual stock prices, the SOSFFNN predicts stock prices that are much closer to the actual stock prices. Hence, SOSFFNN have proven again to have the best predictive performance among the models.

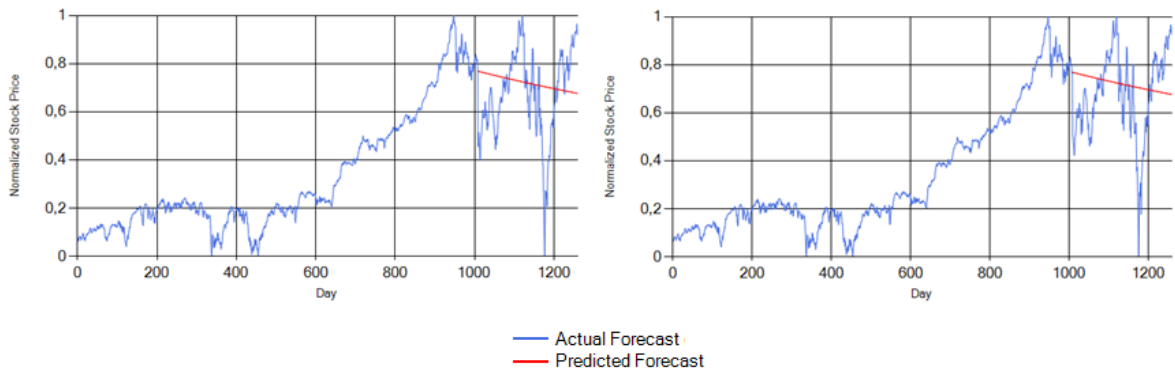

Fig. 19. Graphs displaying the open stock price (left graph) and close stock price (right graph) prediction forecast obtained by the ARIMA model on the Dow Jones Industrial Average Index dataset

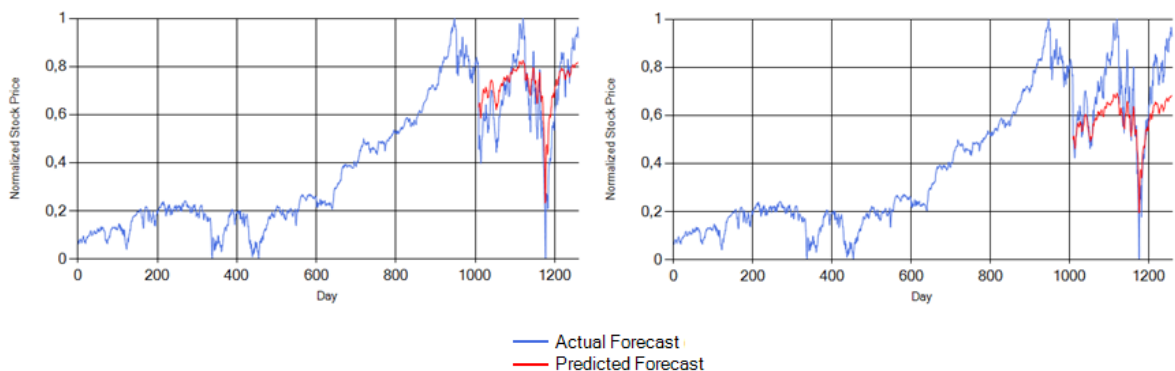

Fig. 20. Graphs displaying the open stock price (left graph) and close stock price (right graph) prediction forecast obtained by the GAFFNN model on the Dow Jones Industrial Average Index dataset

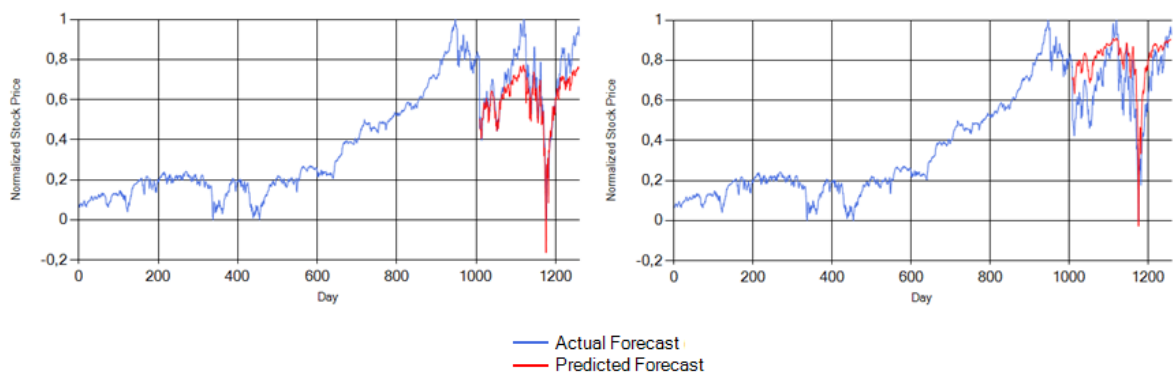

Fig. 21. Graphs displaying the open stock price (left graph) and close stock price (right graph) prediction forecast obtained by the PSOFFNN model on the Dow Jones Industrial Average Index dataset

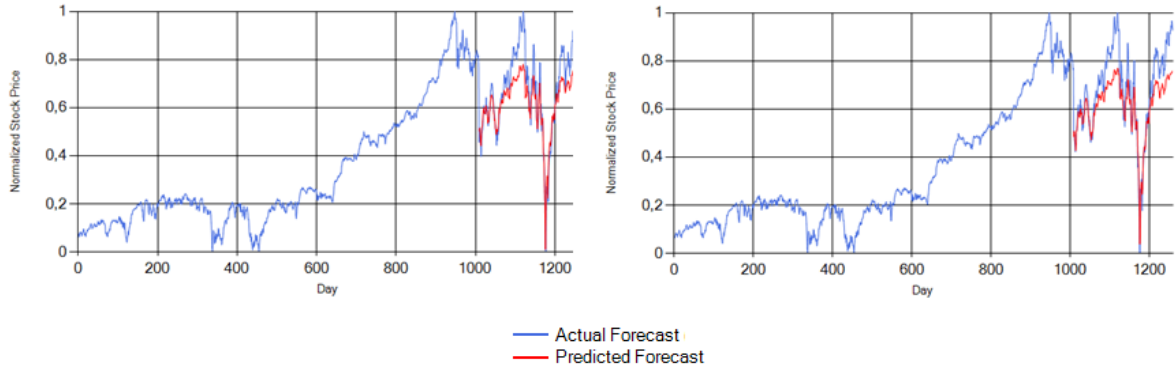

Fig. 22. Graphs displaying the open stock price (left graph) and close stock price (right graph) prediction forecast obtained by the SOSFFNN model on the Dow Jones Industrial Average Index dataset

The NASDAQ Composite index is another stock market index for the United States stock market. The ARIMA, GAFFNN, PSOFNN and the SOSFFNN models were executed on the NASDAQ dataset and the results are shown in Figures 23 to 26, respectively. The ARIMA model displays the worst performance of the models. It over predicts both the opening and closing stock prices for each day. Both predicted forecast graphs in Figure 23 have a slope with a negative gradient with decreasing steepness. The results of the prediction of the opening stock prices are quite similar. Overall, the SOSFFNN predicts stock prices better than do the other models, and in particular the PSOFFNN seems to perform better than the GAFFNN for the prediction of the opening stock prices. The PSOFFNN appears to be the worst amongst the other hybrid models when predicting the closing stock prices. The majority of the predicted closing stock prices are much lower than the actual stock prices. The GAFFNN and SOSFFNN were able to give better predictions and at the same time produce a predicted forecast graph that is similar to the fluctuation trend of the actual stock prices. However, for predicted forecast, the SOSFFNN is the best performing amongst all the models. The predicted closing stock prices for the SOSFFNN are very close to the actual stock prices. Therefore, it can be concluded that the SOSFFNN model is a better forecasting model than are the other models when executed on the various experimented datasets.

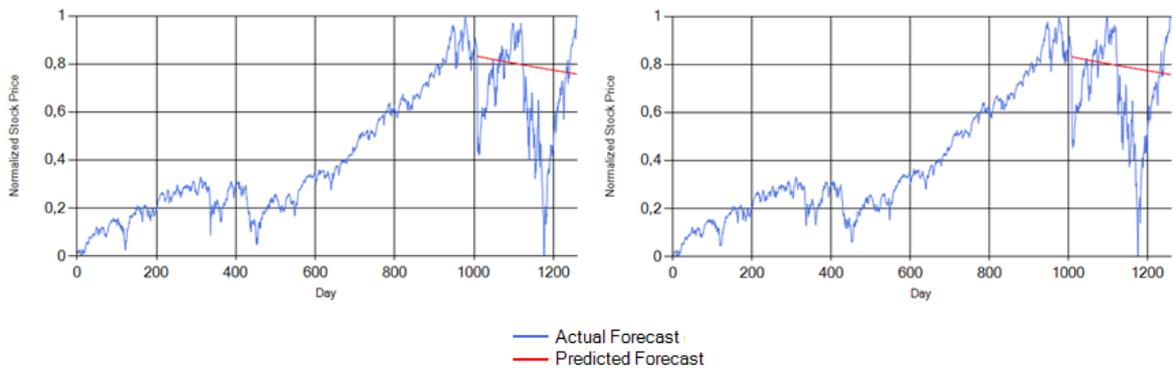

Fig. 23. Graphs displaying the open stock price (left graph) and close stock price (right graph) prediction forecast obtained by the ARIMA model on the NASDAQ Composite dataset

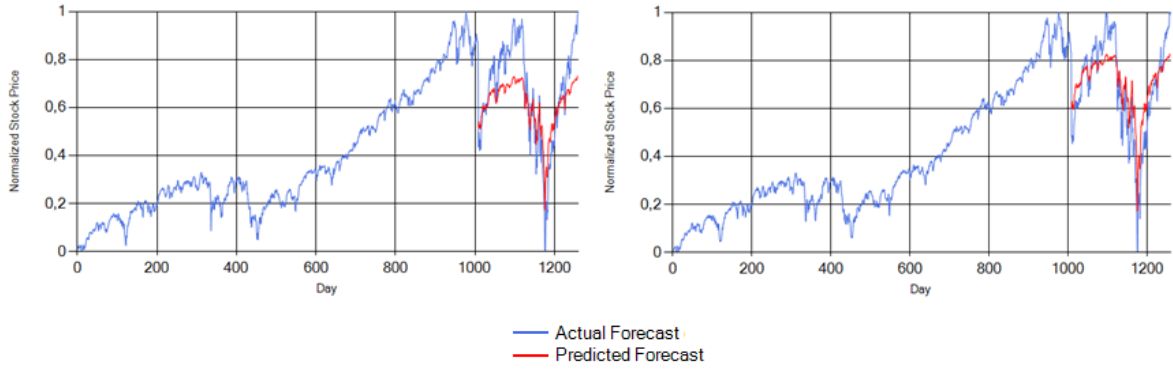

Fig. 24. Graphs displaying the open stock price (left graph) and close stock price (right graph) prediction forecast obtained by the GAFFNN model on the NASDAQ Composite dataset

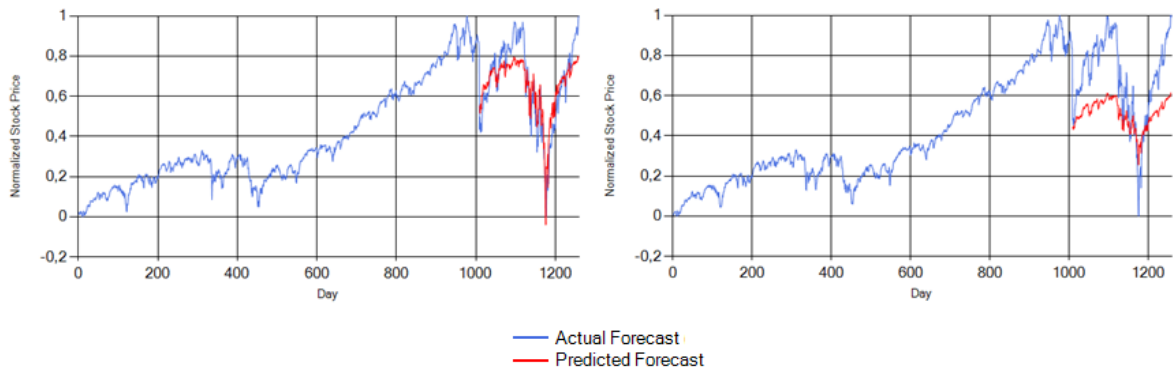

Fig. 25. Graphs displaying the open stock price (left graph) and close stock price (right graph) prediction forecast obtained by the PSOFFNN model on the NASDAQ Composite dataset

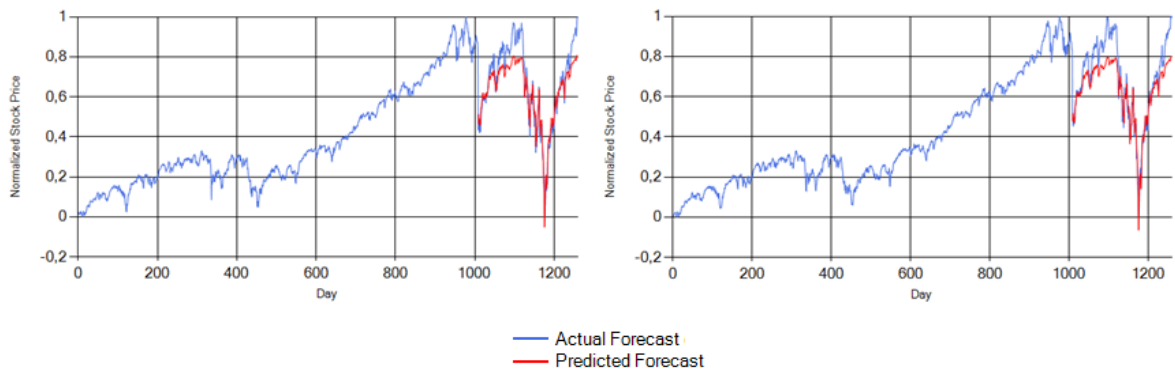

Fig. 26. Graphs displaying the open stock price (left graph) and close stock price (right graph) prediction forecast obtained by the SOSFFNN model on the NASDAQ Composite dataset.

### 4.2. Comparison with literature results

Table 11 shows the results obtained by using the least square support vector machine (LS-SVM) method, which was initially proposed in [34], the least square support vector machine optimized by the PSO algorithm (PSO-LS-SVM) proposed in [35], the artificial neural network trained by backpropagation (NN-BP) proposed in [36] and our proposed SOSFFNN model. Each of the existing models results were compared with the new symbiotic organisms search trained neural network (SOSFFNN) presented in this study. All the algorithms were trained and tested with datasets from January 2009 to January 2012, consisting of eleven stock indices. Each dataset was divided using 70% of the data for training and 30% of the data for testing. These datasets represent different sectors of the market, such as information technology, financial, industrial, health, energy, and materials. The datasets are obtained from Yahoo Finance (https://finance.yahoo.com) and stock market quotes and financial news (https://www.investing.com). The average of 20 executions of the four methods namely PSO-LS-SVM, LS-SVM, NN-BP, and SOSFFNN algorithm are presented for each of the datasets to give a fair comparison. The mean square error was used to evaluate the performance of each of the four models. The lower the mean square error value achieved by the algorithm the better the performance of the algorithm. Therefore, it is obvious from the presented results that the SOSFFNN obtained mean square error values that are significantly better than those achieved by the other models. In fact, the SOSFFNN achieved mean square error values that are very close to zero.

Table 11: Comparison of SOSFFNN with existing models based on Mean Square Error over various datasets.

| Datasets | PSO-LS-SVM | LS-SVM | NN-BP | SOSFFNN |
|---|---|---|---|---|
| Adobe | 0.5317 | 0.5703 | 0.8982 | 0.0056 |
| Oracle | 0.6314 | 0.8829 | 0.9124 | 0.0062 |
| HP | 0.7725 | 1.2537 | 1.9812 | 0.0168 |
| American Express | 0.7905 | 1.0663 | 2.8436 | 0.0135 |
| Bank of New York | 0.4839 | 1.2769 | 1.9438 | 0.0100 |
| Coca-Cola | 0.6823 | 0.9762 | 1.7975 | 0.0079 |
| HoneyWell | 0.9574 | 1.3371 | 2.1853 | 0.0073 |
| Exxon-Mobile | 1.1 | 1.6935 | 2.4891 | 0.0137 |
| AT & T | 0.2911 | 0.4673 | 0.4055 | 0.0107 |
| FMC Corp. | 1.5881 | 2.1034 | 3.5049 | 0.0069 |
| Duke Energy | 0.1735 | 0.6097 | 0.601 | 0.0079 |

## 5. Conclusion

In this paper, the hybrid SOSFFNN model was developed and tested based on the existing hybridization concept of PSO and FFNNs study. Additional comparisons, which involved implementing hybrid GAFFNN and ARIMA model, were carried out to further validate the superior performance of the hybrid SOSFFNN algorithm. The experimental results obtained by the hybrid SOSFFNN revealed that the proposed SOS-trained model outperformed the hybrid PSOFFNN, GAFFNN and ARIMA based models by noticeable margins. However, some of the shortcomings of the proposed SOSFFNN model were identified, and were attributed to the increased implementation complexity given by the combination of two already complex algorithms involved in the hybridization process, coupled with the training challenges of FFNNs model. Future improvements on the proposed SOS trained FFNN could involve training the new model using datasets with higher complexity levels. This could then be adapted to incorporate multi-objective parameters between relative stock prices that may influence another stock's price. Finally, since the proposed model shows promise in the area of stock price prediction, the aspect of the FFNN training complexity may be a useful area of research that would requires greater fine-tuning to achieve a better predictive accuracy. Specifically, a deep experimental investigation on increasing the values of the FFNNs hidden layers may be studied further.

**Conflict of interests**

The authors declare that there is no conflict of interests regarding the publication of the paper.

**References**

[1] Pownall, G., Wasley, C. and Waymire, G., 1993. The stock price effects of alternative types of management earnings forecasts. The Accounting Review, 68(4), p.896.
[2] Pai, P.F. and Lin, C.S., 2005. A hybrid ARIMA and support vector machines model in stock price forecasting. Omega, 33(6), pp.497-505.
[3] Seetharaman, A., Niranjan, I., Patwa, N. and Kejriwal, A., 2017. A study of the factors affecting the choice of investment portfolio by individual investors in Singapore. Accounting and Finance Research, 6(3), p.153.
[4] Montgomery, D.C., Johnson, L.A. and Gardiner, J.S., 1990. Forecasting and time series analysis. McGraw-Hill Companies.

[5] Refenes, A.N., Zapranis, A. and Francis, G., 1994. Stock performance modeling using neural networks: a comparative study with regression models. Neural networks, 7(2), pp.375-388.
[6] Schöneburg, E., 1990. Stock price prediction using neural networks: A project report. Neurocomputing, 2(1), pp.17-27.
[7] White, H., 1988. Economic Prediction Using Neural Networks: The Case of IBM Daily Stock Return. In Proceedings of International Conference on Neural Networks (pp. II-451).
[8] Zhang, G.P., 2003. Time series forecasting using a hybrid ARIMA and neural network model. Neurocomputing, 50, pp.159-175.
[9] Ponnam, L.T., Rao, V.S., Srinivas, K. and Raavi, V., 2016, September. A comparative study on techniques used for prediction of stock market. In 2016 International Conference on Automatic Control and Dynamic Optimization Techniques (ICACDOT) (pp. 1-6). IEEE.
[10] Konno, H. and Yamazaki, H., 1991. Mean-absolute deviation portfolio optimization model and its applications to Tokyo stock market. Management science, 37(5), pp.519-531.
[11] Trippi, R.R., By-Lee, P. and Jae, K., 1995. Artificial intelligence in finance and investing: state-of-the-art technologies for securities selection and portfolio management. McGraw-Hill, Inc.
[12] Wu, H., Zhou, Y., Luo, Q. and Basset, M.A., 2016. Training feedforward neural networks using symbiotic organisms search algorithm. Computational intelligence and neuroscience, 2016.
[13] Adhikari, R. and Agrawal, R.K., 2013. Hybridization of Artificial Neural Network and Particle Swarm Optimization Methods for Time Series Forecasting. International Journal of Applied Evolutionary Computation (IJAEC), 4(3), pp.75-90.
[14] Jabin, S., 2014. Stock market prediction using feed-forward artificial neural network. International Journal of Computer Applications, 99(9), pp.4-8.
[15] Junyou, B., 2007, September. Stock Price forecasting using PSO-trained neural networks. In 2007 IEEE Congress on Evolutionary Computation (pp. 2879-2885). IEEE.
[16] Ezugwu, A.E., Adeleke, O.J., Akinyelu, A.A. and Viriri, S., 2019. A conceptual comparison of several metaheuristic algorithms on continuous optimisation problems. Neural Computing and Applications, pp.1-45.
[17] Ezugwu, A.E. and Prayogo, D., 2018. Symbiotic organisms search algorithm: theory, recent advances and applications. Expert Systems with Applications 119: 184-209.
[18] Ezugwu, A.E., Adeleke, O.J. and Viriri, S., 2018. Symbiotic organisms search algorithm for the unrelated parallel machines scheduling with sequence-dependent setup times. PloS one, 13(7), p.e0200030.
[19] Ezugwu, A.E., 2019. Enhanced symbiotic organisms search algorithm for unrelated parallel machines manufacturing scheduling with setup times. Knowledge-Based Systems, 172, pp.15-32.
[20] Govender, P. and Ezugwu, A.E., 2018. A symbiotic organisms search algorithm for optimal allocation of blood products. IEEE Access, 7, pp.2567-2588.
[21] Govender, P. and Ezugwu, A.E., 2019, January. A Symbiotic Organisms Search Algorithm for Blood Assignment Problem. In International Workshop on Hybrid Metaheuristics (pp. 200-208). Springer, Cham.
[22] Ezugwu, A.E.S., Adewumi, A.O. and Frîncu, M.E., 2017. Simulated annealing based symbiotic organisms search optimization algorithm for traveling salesman problem. Expert Systems with Applications, 77, pp.189-210.
[23] Ezugwu, A.E.S. and Adewumi, A.O., 2017. Discrete symbiotic organisms search algorithm for travelling salesman problem. Expert Systems with Applications, 87, pp.70-78.
[24] Rath, S., Sahu, B.K. and Nayak, M.R., 2019. Application of quasi-oppositional symbiotic organisms search based extreme learning machine for stock market prediction. International Journal of Intelligent Computing and Cybernetics, 12(2), pp.175-193.
[25] Huang, G.B., 2003. Learning capability and storage capacity of two-hidden-layer feedforward networks. IEEE Transactions on Neural Networks, 14(2), pp.274-281.
[26] Huang, G.B., Chen, L. and Siew, C.K., 2006. Universal approximation using incremental constructive feedforward networks with random hidden nodes. IEEE Trans. Neural Networks, 17(4), pp.879-892.
[27] Cadenas, E. and Rivera, W., 2010. Wind speed forecasting in three different regions of Mexico, using a hybrid ARIMA–ANN model. Renewable Energy, 35(12), pp.2732-2738.
[28] Lahmiri, S., 2016. Interest rate next-day variation prediction based on hybrid feedforward neural network, particle swarm optimization, and multiresolution techniques. Physica A: Statistical Mechanics and its Applications, 444, pp.388-396.
[29] Ghanem, W.A.H. and Jantan, A., 2014. Using hybrid artificial bee colony algorithm and particle swarm optimization for training feed-forward neural networks. Journal of theoretical & applied information technology, 67(3).
[30] Panapakidis, I.P. and Dagoumas, A.S., 2017. Day-ahead natural gas demand forecasting based on the combination of wavelet transform and ANFIS/genetic algorithm/neural network model. Energy, 118, pp.231-245.
[31] Fattah, M.A. and Ren, F., 2009. GA, MR, FFNN, PNN and GMM based models for automatic text summarization. Computer Speech & Language, 23(1), pp.126-144.
[32] Pillay, B.J. and Ezugwu, A.E., 2019, July. Stock Price Forecasting Using Symbiotic Organisms Search Trained Neural Networks. In International Conference on Computational Science and Its Applications (pp. 673-688). Springer, Cham..
[33] Nanda, S.J. and Jonwal, N., 2017. Robust nonlinear channel equalization using WNN trained by symbiotic organism search algorithm. Applied Soft Computing, 57, pp.197-209.
[34] Suykens, J.A., De Brabanter, J., Lukas, L. and Vandewalle, J., 2002. Weighted least squares support vector machines: robustness and sparse approximation. Neurocomputing, 48(1-4), pp.85-105.
[35] Liao, R., Zheng, H., Grzybowski, S. and Yang, L., 2011. Particle swarm optimization-least squares support vector regression based forecasting model on dissolved gases in oil-filled power transformers. Electric Power Systems Research, 81(12), pp.2074-2080.
[36] Wang, J.Z., Wang, J.J., Zhang, Z.G. and Guo, S.P., 2011. Forecasting stock indices with back propagation neural network. Expert Systems with Applications, 38(11), pp.14346-14355.

[37] Yaghini, M., Khoshraftar, M.M. and Fallahi, M., 2013. A hybrid algorithm for artificial neural network training. Engineering Applications of Artificial Intelligence, 26(1), pp.293-301.

[38] Socha, K., Blum, C., 2007. An ant colony optimization algorithm for continuous optimization: application to feed-forward neural network training. Neural Comput. Appl. 16 (May (3)), 235–247.

[39] Sarangi, P.P., Sahu, A., Panda, M., 2014. Training a feed-forward neural network using artificial bee colony with back-propagation algorithm. In: Proceedings of the Intelligent Computing, Networking, and Informatics. Springer, pp. 511–519.

[40] Kulluk, S., Ozbakir, L., Baykasoglu, A., 2012. Training neural networks with harmony search algorithms for classification problems. Eng. Appl. Artif. Intell. 25 (February (1)), 11–19.

[41] Horng, M.-H., Lee, M.-C., Liou, R.-J., Lee, Y.-X., 2012. Firefly meta-heuristic algorithm for training the radial basis function network for data classification and disease diagnosis. In: Parpinelli, R., Lopes, H.S. (Eds.), Theory and New Applications of Swarm Intelligence, InTech.

[42] Vázquez, R.A., 2011. Training spiking neural models using cuckoo search algorithm. In: Proceedings of the IEEE Congress Evolutionary Computation (CEC), 2011, pp. 679–686.

[43] Ghalambaz, M., Noghrehabadi, A., Behrang, M., Assareh, E., Ghanbarzadeh, A., Hedayat, N., 2011. A hybrid neural network and gravitational search algorithm (HNNGSA) method to solve well known wessinger's equation. World Acad. Sci. Eng. Technol. 5, 803–807.

[44] Ulagammai, M., Venkatesh, P., Kannan, P., Padhy, N.P., 2007. Application of bacterial foraging technique trained artificial and wavelet neural networks in load forecasting. Neurocomputing 70 (16), 2659–2667.

[45] Ghasemiyeh, R., Moghdani, R. and Sana, S.S., 2017. A hybrid artificial neural network with metaheuristic algorithms for predicting stock price. Cybernetics and Systems, 48(4), pp.365-392.

[46] Göçken, M., Özçalıcı, M., Boru, A. and Dosdoğru, A.T., 2016. Integrating metaheuristics and artificial neural networks for improved stock price prediction. Expert Systems with Applications, 44, pp.320-331.

[47] Ezugwu, A.E., Otegbeye, O., Govender, P. and Odo, J., 2020. Computational Intelligence Approach to Dynamic Blood Allocation with ABO-Rhesus Factor Compatibility under Real-world Scenario. IEEE Access, vol. 8, pp. 97576-97603, 2020, doi: 10.1109/ACCESS.2020.2997299.

[48] Rajah, V. and Ezugwu, A.E., 2020, March. Hybrid Symbiotic Organism Search algorithms for Automatic Data Clustering. In 2020 Conference on Information Communications Technology and Society (ICTAS) (pp. 1-9). IEEE.

[49] Ezugwu, A.E., Olusanya, M.O. and Govender, P., 2019. Mathematical model formulation and hybrid metaheuristic optimization approach for near-optimal blood assignment in a blood bank system. Expert Systems with Applications, 137, pp.74-99.

[50] Ezugwu, A.E. and Adewumi, A.O., 2017. Soft sets based symbiotic organisms search algorithm for resource discovery in cloud computing environment. Future Generation Computer Systems, 76, pp. 33-50.